\documentclass[twocolumn, twocolappendix]{aastex701}

\usepackage{amsmath}

\def\kms{km~s$^{-1}$}
\def\S{s$^{-1}$}
\def\Jybeam{Jy~beam$^{-1}$}

\def\HI{H\,{\small{I}}}

\def\degree{$^{\circ}$}
\def\deg2{deg$^{2}$}
\def\HIFAST{HiFAST}

\def\HD2{HD$^{2}$}

\def\SOFIA{SoFiA}
\def\Msun{$\mathrm{M_{\odot}}$}

\defcitealias{Springob2005}{Sp05}
\defcitealias{Yu2020}{Yu20}

\shorttitle{The \HD2\ Pilot Survey}
\shortauthors{Xu et al.}

\received{Feb 14, 2026}
\revised{April 15, 2026}
\revised{May 10, 2026}
\revised{May 17, 2026}
\accepted{May 19, 2026}

\submitjournal{ApJS}

\begin{document}

\title{The FAST Hundred-Deg$^2$ \HI\ Deep (HD$^2$) Survey: Early Results from the Pilot Survey}

\correspondingauthor{Jie Wang, Chen Xu, Yingjie Jing}

\author[0000-0003-0062-4705]{Chen Xu}
\affiliation{National Astronomical Observatories, Chinese Academy of Sciences, Beijing 100101, China}
\affiliation{School of Astronomy and Space Science, University of Chinese Academy of Sciences, Beijing 100049, China}
\email[show]{xuchen[at]nao.cas.cn}

\author{Yingjie Jing}
\altaffiliation{FAST fellow}
\affiliation{National Astronomical Observatories, Chinese Academy of Sciences, Beijing 100101, China}
\email[show]{jyj[at]nao.cas.cn}

\author[0000-0002-9937-2351]{Jie Wang}
\affiliation{National Astronomical Observatories, Chinese Academy of Sciences, Beijing 100101, China}
\affiliation{Institute for Frontiers in Astronomy and Astrophysics, Beijing Normal University, Beijing 102206, China}
\affiliation{School of Astronomy and Space Science, University of Chinese Academy of Sciences, Beijing 100049, China}
\email[show]{jie.wang[at]nao.cas.cn}

\author{Hu Zou}
\affiliation{National Astronomical Observatories, Chinese Academy of Sciences, Beijing 100101, China}
\affiliation{School of Astronomy and Space Science, University of Chinese Academy of Sciences, Beijing 100049, China}
\email{zouhu[at]nao.cas.cn}

\author{Wei Du}
\affiliation{National Astronomical Observatories, Chinese Academy of Sciences, Beijing 100101, China}
\email{wdu[at]nao.cas.cn}

\author[0000-0002-9066-370X]{Niankun Yu}
\affiliation{Max Planck Institute for Radio Astronomy, Auf dem Hügel 69, 53121 Bonn, Germany}
\email{nyu[at]mpifr-bonn.mpg.de}

\author[0000-0003-4936-8247]{Hong Guo}
\affiliation{Shanghai Astronomical Observatory, Chinese Academy of Sciences, Shanghai, 200030, China}
\email{guohong[at]shao.ac.cn}

\author{Zerui Liu}
\affiliation{National Astronomical Observatories, Chinese Academy of Sciences, Beijing 100101, China}
\affiliation{School of Astronomy and Space Science, University of Chinese Academy of Sciences, Beijing 100049, China}
\email{liuzr[at]bao.ac.cn}

\author[0009-0009-1428-3375]{Qingze Chen}
\affiliation{National Astronomical Observatories, Chinese Academy of Sciences, Beijing 100101, China}
\affiliation{School of Astronomy and Space Science, University of Chinese Academy of Sciences, Beijing 100049, China}
\email{chenqz[at]bao.ac.cn}

\author[0009-0006-2730-7020]{Tiantian Liang}
\affiliation{Department of Mathematics and Physics, North China Electric Power University, Baoding 071003, China}
\affiliation{Hebei Key Laboratory of Physics and Energy Technology, North China Electric Power University, Baoding 071003, China}
\email{ttliang[at]nao.cas.cn}

\author[0009-0008-4862-2219]{Zhipeng Hou}
\affiliation{National Astronomical Observatories, Chinese Academy of Sciences, Beijing 100101, China}
\affiliation{School of Astronomy and Space Science, University of Chinese Academy of Sciences, Beijing 100049, China}
\email{houzp[at]bao.ac.cn}

\author[0009-0002-2423-4613]{Yiwei Xu}
\affiliation{National Astronomical Observatories, Chinese Academy of Sciences, Beijing 100101, China}
\affiliation{School of Astronomy and Space Science, University of Chinese Academy of Sciences, Beijing 100049, China}
\email{xuyw[at]bao.ac.cn}

\author[0000-0002-2884-9781]{Xiao Li}
\affiliation{Department of Astronomy, Tsinghua University, Beijing 100084, China}
\affiliation{Max Planck Institute for Radio Astronomy, Auf dem Hügel 69, 53121 Bonn, Germany}
\email{xli[at]mpifr-bonn.mpg.de}

\author[0000-0002-1908-0384]{Huijie Hu}
\affiliation{Institute of Astronomy and Astrophysics, Anqing Normal University, Anqing 246133, China}
\affiliation{National Astronomical Observatories, Chinese Academy of Sciences, Beijing 100101, China}
\email{huhuijie[at]nao.cas.cn}

\author[0009-0000-9757-450X]{Ziming Liu}
\affiliation{National Astronomical Observatories, Chinese Academy of Sciences, Beijing 100101, China}
\email{zmliu[at]nao.cas.cn}

\author[0000-0002-4328-538X]{Pinsong Zhao}
\affiliation{Kavli Institute for Astronomy and Astrophysics, Peking University, Beijing 100871, China}
\affiliation{National Astronomical Observatories, Chinese Academy of Sciences, Beijing 100101, China}
\email{pszhao1993[at]pku.edu.cn}

\author[0000-0002-2853-3808]{Taotao Fang}
\affiliation{Department of Astronomy, Xiamen University, Xiamen, Fujian 361005, China}
\email{fangt[at]xmu.edu.cn}

\author[0009-0006-3885-9728]{Liang Gao}
\affiliation{Institute for Frontiers in Astronomy and Astrophysics, Beijing Normal University, Beijing 102206, China}
\affiliation{School of Physics and Astronomy, Beijing Normal University, Beijing 100875, China}
\affiliation{School of Physics and Laboratory of Zhongyuan Light, Zhengzhou University, Zhengzhou 450001, China}
\email{lgao[at]bnu.edu.cn}

\author[0000-0002-7972-3310]{Qi Guo}
\affiliation{Institute for Frontiers in Astronomy and Astrophysics, Beijing Normal University, Beijing 102206, China}
\affiliation{School of Physics and Astronomy, Beijing Normal University, Beijing 100875, China}
\email{qguo[at]bnu.edu.cn}

\author[0000-0002-3890-3729]{Qiusheng Gu}
\affiliation{School of Astronomy and Space Science, Nanjing University, Nanjing 210093, China}
\email{qsgu[at]nju.edu.cn}

\author[0000-0003-2405-5930]{Zhen Jiang}
\affiliation{National Astronomical Observatories, Chinese Academy of Sciences, Beijing 100101, China}
\email{jiangzhen[at]nao.cas.cn}

\author[0000-0002-5458-4254]{Xi Kang}
\affiliation{Institute for Astronomy, School of Physics, Zhejiang University, Hangzhou 310027, China}
\email{kangxi[at]zju.edu.cn}

\author[0000-0002-7660-2273]{Xu Kong}
\affiliation{Department of Astronomy, University of Science and Technology of China, Hefei 230026, China}
\affiliation{School of Astronomy and Space Sciences, University of Science and Technology of China, Hefei 230026, China}
\email{xkong[at]ustc.edu.cn}

\author[0000-0002-8711-8970]{Cheng Li}
\affiliation{Department of Astronomy, Tsinghua University, Beijing 100084, China}
\email{cli2015[at]mail.tsinghua.edu.cn}

\author[0000-0002-4025-7877]{Jun Pan}
\affiliation{Chinese Academy of Sciences South America Center for Astronomy, National Astronomical Observatories, CAS, Beijing 100101, China}
\email{jpan[at]bao.ac.cn}

\author[0000-0002-2504-2421]{Tao Wang}
\affiliation{School of Astronomy and Space Science, Nanjing University, Nanjing 210093, China}
\email{taowang[at]nju.edu.cn}

\author{Yanrui Zhou}
\affiliation{Department of Astronomy, School of Physics and Astronomy, Shanghai Jiao Tong University, Shanghai 200240, China}
\affiliation{State Key Laboratory of Dark Matter Physics, School of Physics and Astronomy, Shanghai Jiao Tong University, Shanghai 200240, China}
\email{123zyr[at]sjtu.edu.cn}

\author[0000-0002-5762-7571]{Wenting Wang}
\affiliation{State Key Laboratory of Dark Matter Physics, School of Physics and Astronomy, Shanghai Jiao Tong University, Shanghai 200240,
China}
\affiliation{Key Laboratory for Particle Astrophysics and Cosmology (MOE)/Shanghai Key Laboratory for Particle Physics and Cosmology,
Shanghai 200240, China}
\email{wenting.wang[at]sjtu.edu.cn}

\author[0000-0002-6593-8820]{Jing Wang}
\affiliation{Kavli Institute for Astronomy and Astrophysics, Peking University, Beijing 100871, China}
\email{jwang_astro[at]pku.edu.cn}

\begin{abstract}

The Hundred-deg$^2$ \HI\ Deep (\HD2) survey carried out with the Five-hundred-meter Aperture Spherical Telescope (FAST) is planned to map a contiguous region within the DESI DR1 footprint, achieving an effective integration time of 20 minutes for each pointing and a uniform detection sensitivity of 0.28~m\Jybeam\ at 4.8~\kms\ resolution. 
We present early results from the pilot \HD2\ survey: a 10-\deg2\ field overlapping with HSC-SSP and the DESI EDR SV3, observed with an integration time of 7.3 minutes per beam and the rms of 0.45~m\Jybeam\ at 4.8~\kms\ resolution. 
We identify 339 \HI\ sources at $z<0.09$, corresponding to $\sim$34 detections per \deg2, nearly six times higher than the detection rate of the wide-field surveys.  
Optical counterparts are primarily identified using DESI redshifts, yielding a matching rate and correctness exceeding 90\% for galaxies with $r<19.5$ mag, a substantial improvement over SDSS.  
Under the constraint of $r < 17.8$ mag and $0.01 < z < 0.05$, nearly 50\% of galaxies in the DESI BGS samples have \HI\ detections in this pilot survey. The optical properties of these \HI-detected galaxies span nearly the entire parameter range of the DESI sample. 
The gas fraction scaling relations versus stellar mass, stellar mass surface density, $\mathrm{NUV}-r$, and specific star formation rate are consistent with previous surveys, e.g., ALFALFA, DINGO, and xGASS.
These results justify the feasibility of the full \HD2\ survey, which will build a high-completeness \HI\ census over a contiguous area to probe the cold gas scaling relations of galaxies over different scales.

\end{abstract}


\section{Introduction} \label{sec:intro}

Cold gas fuels star formation, with neutral atomic hydrogen (\HI) and molecular hydrogen providing key insights into galaxy evolution. 
\HI\ concentrated in the optical disk fuels star formation \citep[e.g., ][]{Wang2020, Yu2022, Lee2025} and serves as a crucial reservoir in the baryonic cycle, linking the dense molecular gas and hot ionized gas. 
Measurements of the \HI\ 21 cm line flux and spectra provide key insights into the atomic gas content and spatial distribution of galaxies \citep[e.g., ][]{Yu2022a, Zuo2022, Lin2025}. Neutral hydrogen also serves as a sensitive tracer of galaxy interactions, environmental effects, and large-scale structure.
A census of \HI\ 21 cm emission is thus essential for a comprehensive picture of galaxy formation and evolution.

Since the early 21st century, blind extragalactic \HI\ surveys have assembled large samples in the local Universe \citep[e.g.][]{Barnes2001, Giovanelli2005, Li2018, Zhang2024, Koribalski2020}, covering thousands of square degrees but limited in redshift and sensitivity. In contrast, surveys targeting higher redshifts \citep[e.g.][]{Jaffe2013, Hoppmann2015, Blyth2018, Hess2019, Xi2022} can detect \HI\ up to $z \sim 0.4$ \citep{Xi2024}, but require hundreds of hours for only a few square degrees, making them susceptible to cosmic variance.
Medium-area deep surveys provide a balance between coverage and sensitivity. 
The Deep Investigation of Neutral Gas Origins survey \citep[DINGO, ][]{Meyer2010, Rhee2022} using ASKAP\footnote{The Australian Square Kilometre Array Pathfinder.}, will cover about 60 -- 150 \deg2, while the \HI\ project of the MeerKAT International GHz Tiered Extragalactic Exploration survey \citep[MIGHTEE-HI, ][]{Maddox2021} and the MeerKAT Fornax Survey \citep{Serra2023} will probe $\sim$32 \deg2 deep fields.

However, those medium-area deep surveys utilizing interferometry often lack the sensitivity of single-dish telescopes, particularly when compared with the Five-hundred-meter Aperture Spherical Radio Telescope \citep[FAST,][]{Nan2011, Jiang2019, Jiang2020}. 
FAST's large effective aperture, multibeam receiver, and low system temperature make it well-suited for deep, medium-area \HI\ surveys.
Leveraging this capability, we initiated the Hundred-\deg2 \HI\ Deep (\HD2) survey, designed to cover $\sim$100 \deg2\ with higher sensitivity than wide-field \HI\ surveys. As the first step, we conducted a 10 \deg2 pilot survey reported here. 
The \HD2\ survey aims to achieve a sensitivity that is two orders of magnitude deeper than those interferometric surveys, and deliver an \HI\ source catalog with higher source density.
It can extend the detection limit toward low-mass systems, enabling a more accurate determination of the faint end of the \HI\ mass function in the local Universe, and then to compare observations with simulations and models \citep[e.g. ][]{Martin2010, Jones2018, Guo2023, Ma2025}.

It should be noted that blind \HI\ surveys often miss \HI-poor galaxies, especially red and early-type ones, leading to biases in \HI\ mass relations derived from \HI-rich samples. To mitigate this, the \textit{GALEX} Arecibo SDSS Survey \citep[GASS,][]{Catinella2010} combines Arecibo Legacy Fast Arecibo $L$-band Feed Array survey \citep[ALFALFA, ][]{Giovanelli2005, Haynes2011, Haynes2018} with deeper optical-targeted observations, later extended to xGASS \citep{Catinella2018}, which is mainly stellar-mass selected and \HI\ gas-fraction ($M_\mathrm{\HI} / M_\star$) limited.
Another approach is spectral stacking, widely applied in \HI\ surveys with optical catalogs to probe scaling relations, environmental effects \citep[e.g.][]{Fabello2011, Brown2015, Brown2017, Sinigaglia2022, Rhee2022}, and the cosmic \HI\ density $\Omega_{\mathrm{HI}}$ \citep[e.g.][]{Delhaize2013, Rhee2013, Rhee2022, Chowdhury2020, Hu2020}. 
Beyond spectral stacking of galaxies, stacking within halos enables studies of the \HI\ -- halo mass relation against simulations. Here, we adopt stacking for both galaxies and groups to explore scaling relations at low \HI\ fractions.

For a deep \HI\ survey, comparable optical imaging and extensive redshift coverage are essential to identify optical counterparts (OCs), perform spectral stacking, and trace large-scale structures.
With the early data release (EDR) and first data release (DR1) from the Dark Energy Spectroscopic Instrument \citep[DESI,][]{DESICollaboration2024, DESICollaboration2025}, we now have the opportunity to evaluate DESI's performance against the Sloan Digital Sky Survey \citep[SDSS,][]{York2000, Almeida2023} in the context of radio surveys.
We further combine deep photometry from the Hyper Suprime-Cam Subaru Strategic Program \citep[HSC-SSP;][]{Aihara2018, Aihara2022} and the DESI Legacy Imaging Surveys \citep{Dey2019} to identify OCs for \HI\ detections and derive some of the galaxy properties. These resources also enable the \HD2\ survey to uncover rare populations, such as low surface brightness galaxies \citep[e.g.][]{McGaugh1997, Du2024, OBeirne2025}, ultra-diffuse galaxies \citep[e.g.][]{VanDokkum2015, Hu2023}, \HI-rich early-type galaxies \citep{DiSeregoAlighieri2007, Li2024}, and dark systems like RELHICs \citep[REionization-Limited \HI\ Clouds,][]{Benitez-Llambay2017, Zhou2023}.

The structure of this work is as follows. 
In Section~\ref{sec:obs}, we introduce the target field and scan strategy of the \HD2\ pilot survey.
Section~\ref{sec:data_reduction} describes the data reduction process.
We present the \HI\ extragalactic source catalog and compare the source density and detection rate with other surveys in Section~\ref{sec:result}.
Section~\ref{sec:scaling} presents the \HI\ mass contents and scaling relations using direct detections and spectral stacking,
followed by our summary in Section~\ref{sec:summary}.
In this paper, we adopt a $\rm{\Lambda}$CDM cosmology with $\Omega_m =  0.3$, $\Omega_{\Lambda} = 0.7$, and Hubble constant value $H_0 = 70~ \mathrm{km~s^{-1}~Mpc^{-1}}$.

\section{Observations}\label{sec:obs}

FAST is equipped with a 19-beam receiver \citep{Jiang2020}, enabling efficient multi-beam surveys. The central beam has a Gaussian-like profile with a half-power beamwidth of 2\farcm 9 at 1.42 GHz.
During observations, the system temperature ranges from 19 to 23 K, and the gain varies between 14 and 16 K Jy$^{-1}$ \citep{Liu2024}. We employ a wide-band spectral-line backend with a resolution of 7.63 kHz (1.61~\kms\ at $z=0$), recording two linear polarizations, \texttt{XX} and \texttt{YY}, separately.
The backend can cover the $L$-band range of 1.0 -- 1.5 GHz; however, strong radio frequency interference (RFI) between 1.15 and 1.3 GHz limits our initial analysis to $z < 0.09$ \citep{Zhang2022}.
The final data cube is gridded with $1'$ pixels, yielding a map median rms of 0.45~m\Jybeam\ at 4.8 \kms, and a median spectral rms of $\sim$0.46 mJy at 10~\kms\ resolution from the spatially
integrated line profiles. Those observation details are summarized in Table~\ref{tab:survey}.

\begin{deluxetable}{ll}[h!]
\tabletypesize{\scriptsize}
\tablewidth{1pt} 
\tablecaption{The \HD2\ pilot Survey Technical Details}\label{tab:survey}
\tablehead{\colhead{Parameters} & \colhead{Details}}
\startdata 
R.A. range & $15^{\mathrm{h}}36^{\mathrm{m}}00^{\mathrm{s}} \text{ -- } 16^{\mathrm{h}}32^{\mathrm{m}}00^{\mathrm{s}}$ \\
Decl. range & $42^\circ 48' 00'' \text{ -- } 43^\circ 48' 00''$ \\
Number of beams & 19 \\
Beam size & 2\farcm 9 at 1.42 GHz \\
Polarizations & 2 (\tt{XX, YY}) \\
Gain & 14 -- 16 K Jy$^{-1}$ \\
$T_\mathrm{sys}$ & 19 --23 K \\
Full frequency range & 1000 -- 1500 MHz \\
Full spectral channels & 65,536 \\
Processed frequency range & 1300 -- 1440 MHz \\
\SOFIA\ searching frequency range & 1305 -- 1419 MHz \\
Redshift range & 0.001 -- 0.09 \\
Velocity range & 300 -- 26,600~\kms\ \\
Bandwidth (low-z used) & 140 MHz \\
Channel resolution & 7.6 kHz (W band), \\
 & 22.8 kHz (after down-sample) \\
Spectral resolution (at $z=0$) & 1.6~\kms\ (W band), \\
 & 4.8~\kms\  (after down-sample), \\
 & 10~\kms\ (after Hann smooth) \\
Map median rms & 0.45~m\Jybeam\ at 4.8~\kms\ \\
Spectral median rms & 0.46 mJy at 10~\kms\ \\
Grid pixel size & 1$'$ \\
\enddata
\end{deluxetable}

\subsection{Field Selection}

When the zenith angle (ZA) is below 26.5\degree, FAST achieves optimal aperture efficiency and system temperature \citep{Jiang2020}.
Therefore, we selected a strip at a declination of $\sim$43\degree, where the minimum ZA is 18\degree. This region also suffers less RFI than equatorial fields, which are more strongly affected by geostationary satellites.

\begin{figure*}[ht!]
\plotone{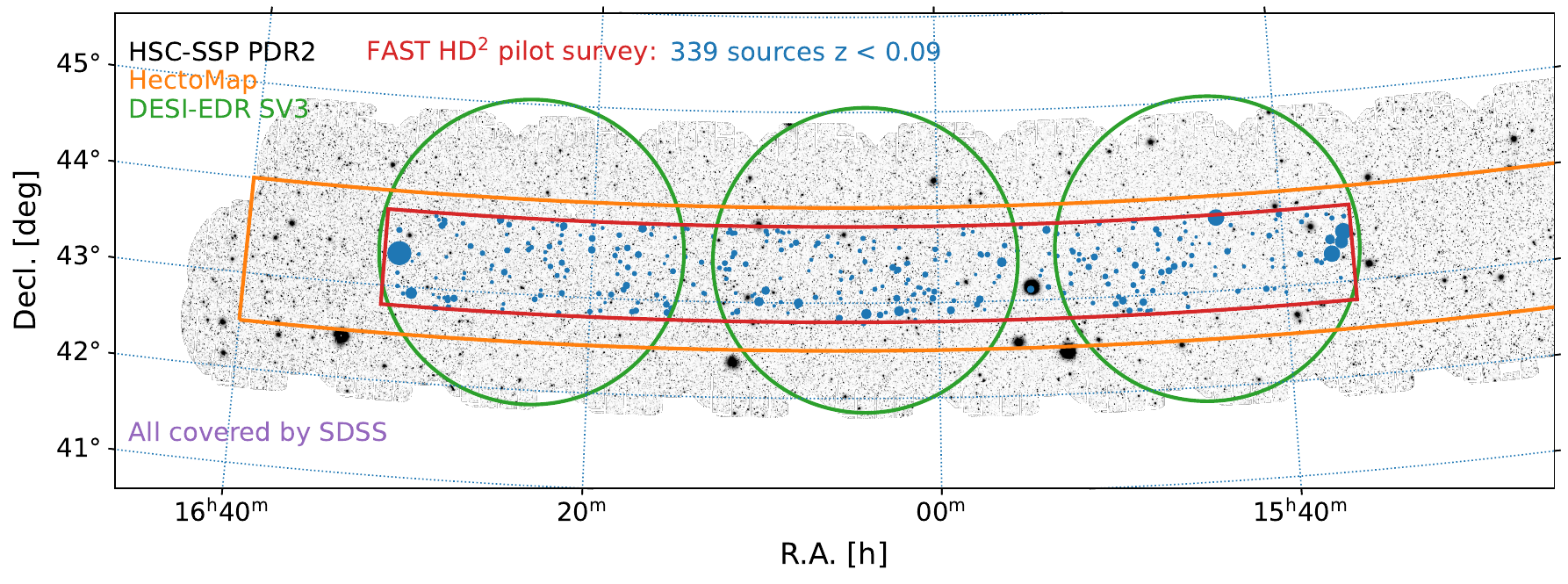}
\caption{The sky distribution of the pilot survey. The background image is from the HSC-SSP PDR2 wide layer data \citep{Aihara2019}. 
The redshift survey named HectoMAP \citep{Sohn2023} covers the orange strip across 43.25\degree. 
Three green circles show the footprint of DESI EDR SV3 \citep{DESICollaboration2024} spectra distribution in this area. 
The entire region is covered by SDSS. 
The red box is our 10-\deg2 target field for FAST observation. 
We detect 339 sources at $z<0.09$ plotted as blue dots, and the dot size indicates the \HI\ flux.}
\label{fig:area}
\end{figure*}

The pilot survey area is covered by three spectroscopic surveys: SDSS \citep{York2000}, HectoMAP \citep{Sohn2023}, and DESI EDR \citep{DESICollaboration2024}, as well as the deep imaging survey HSC-SSP PDR2 (wide) \citep{Aihara2019}. The overlap of our observations and optical surveys is shown in Figure~\ref{fig:area}. HectoMAP provides spectroscopy for a 1.5\degree-wide strip, while HSC-SSP has imaged the same region with its wide survey mode. In 2023, DESI released high-completeness early spectra from the One-Percent Survey (Survey Validation 3, SV3), covering three tiles overlapping with the HSC-SSP fields and achieving the highest completeness. A total of 339 \HI\ detections at $z<0.09$ are also marked in Figure~\ref{fig:area}.

\subsection{Scan Strategy}

Since the pilot area spans only 1 hour in right ascension (R.A.), daily observations in drift scan mode for a single hour are not efficient. Therefore, we selected the Multibeam On-The-Fly (OTF) mapping mode, which allows for lower scan speeds compared to the drift scan. We employed a scan speed of $5''~\mathrm{s}^{-1}$, which is approximately one-third of the drift scan speed. 
Our targeted 10-\deg2 field was mapped with three independent OTF scans. Each scan was repeated over three nights, leading to a total of 9 nights of observation from late July to early August 2023 (PI: Jie Wang; project ID: ZD2022\_4, ZD2023\_4). This resulted in a total of 25.2 hours integration time (approximately 2.8 hours per night), and an effective integration time equivalent to 9 passes in drift scan mode, corresponding to 7.3 minutes per beam solid angle.
The final \HD2\ survey will cover approximately 100 \deg2\ with at least 24 passes of equal drift scan, resulting in a cumulative integration time of about 20 minutes per beam.

\section{Data Reduction} \label{sec:data_reduction}

\subsection{Calibration and Imaging}\label{sec:cali}

\begin{figure*}[ht!]
\plotone{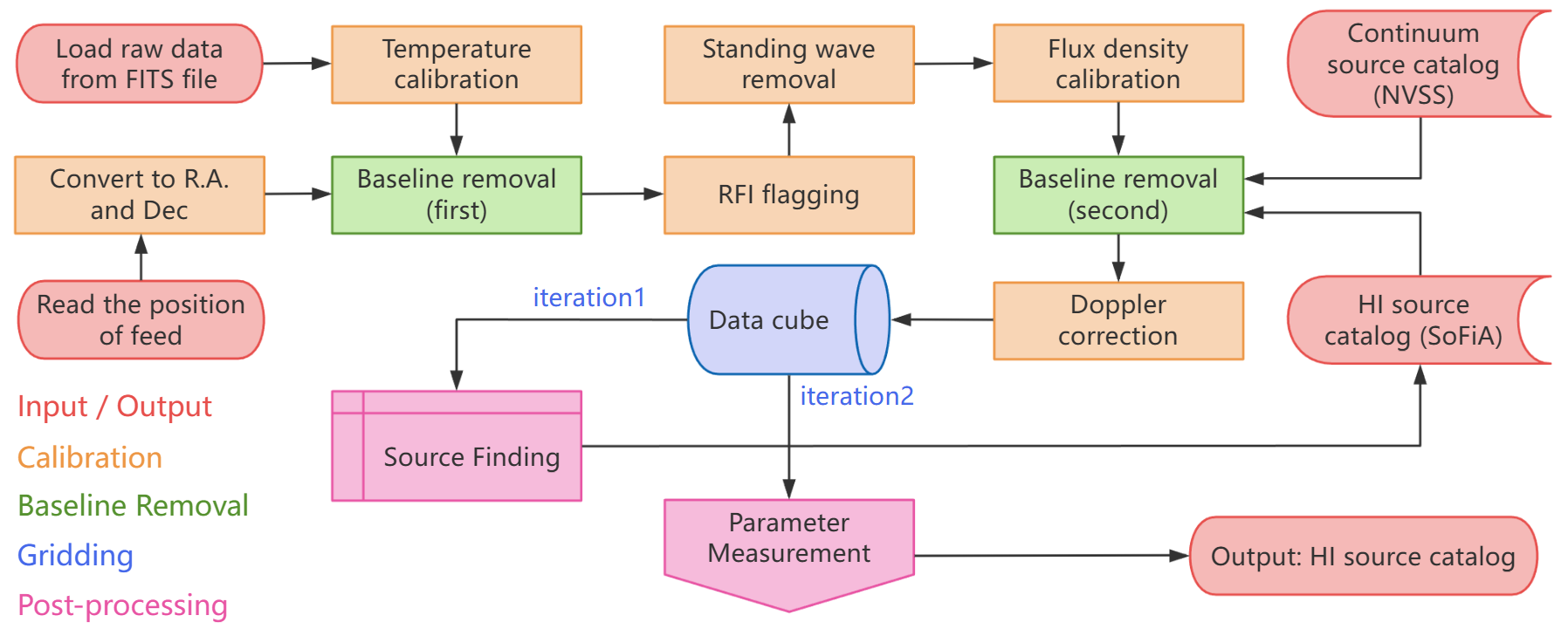}
\caption{Flowchart of the \HIFAST\ calibration and imaging process for point source surveys.}
\label{fig:hifast_pipe}
\end{figure*}

We use \HIFAST\footnote{\url{https://hifast.readthedocs.io}}, a dedicated pipeline for FAST \HI\ data calibration and imaging \citep{Jing2024}, to reduce the raw spectra into spectral cubes.
\HIFAST\ is a modular and flexible pipeline, comprising modules for noise diode calibration, baseline fitting, standing-wave removal using an FFT-based method \citep{Xu2025}, flux density calibration \citep{Liu2024}, Doppler correction, stray radiation correction \citep{Chen2026}, and gridding to produce data cubes.
The data reduction workflow of \HIFAST\ for point sources is shown in Figure~\ref{fig:hifast_pipe}.

The temperature calibration follows the procedure described in \citet{Jing2024}. The spectra were sampled with a time resolution of 0.5 s, and the noise diode injected a high-power signal for 2 s every 5 min.
For the flux calibration, we did not use an external calibrator but instead adopted the gain curve from \citet{Jiang2019}.
Although we discarded some portions of the data affected by RFI during the three-pass observations, all regions were still covered in at least one pass.
Artifacts related to 0.5-MHz harmonic RFI \citep{Xu2025} were evident and difficult to remove, particularly in beams M04 and M05.
\citet{Xu2025} reported three types of standing waves at FAST, with frequency periods of 1.09, 1.92, and 0.039 MHz. We mitigate these by applying an FFT-based filtering method to remove the standing-wave modes in Fourier space, which is effective for time-varying signals.

Besides the standing waves and source emission or absorption lines, we subtract other components such as the sky background, ground radiation, continuum emission, and other contributions in the system temperature. This requires baseline fitting and bandpass correction. For the point-source survey, we fit and remove a low-order polynomial baseline from each spectrum and then apply the `MedMed' method to correct frequency-dependent bandpass variations \citep{Jing2024}.
To avoid underestimating source fluxes, we perform a second iteration of baseline fitting and cube imaging after source finding, using \HI\ candidates and continuum survey catalog as input in Figure~\ref{fig:hifast_pipe}. 

\subsection{Source Finding}\label{sec:source_finding}

We identified \HI\ sources through a multi-stage process combining automated detection with \SOFIA\ 2\footnote{HI Source Finding Application version 2: \url{https://gitlab.com/SoFiA-Admin/SoFiA-2}} and subsequent visual inspection to ensure catalog reliability. \SOFIA\ 2 is a source finder for 3D spectral-line data, widely used in ongoing \HI\ surveys \citep{Serra2015, Westmeier2022}, including the Wide-field ASKAP $L$-band Legacy All-sky Blind Survey \citep[WALLABY, ][]{Westmeier2022}, DINGO \citep{Rhee2022}, and the FAST All Sky \HI\ Survey \citep[FASHI, ][]{Zhang2024}.
To minimize the risk of missing real sources, we adopted low parameter thresholds in the initial search with the smooth-and-clip (S+C) finder. 
The S+C finder was run with spatial smoothing kernels of \texttt{scfind.kernelsXY = [0, 3, 4]} pixels and spectral smoothing kernels of \texttt{scfind.kernelsZ = [0, 3, 7, 15]} channels, with a pixel-level detection threshold of \texttt{scfind.threshold = 4.0} times the smoothed RMS noise. 
This approach successfully recovered faint candidates but also produced many false detections (e.g., RFI, baseline artifacts). Therefore, careful human verification is required after the linking and reliability filtering stages.

During manual verification, we examined the morphology and consistency of candidates across two polarizations and three independent scans. We inspected their zeroth-moment maps and spectra to distinguish true \HI\ signals from artifacts. 
For uncertain cases, we used \texttt{ipyaladin}\footnote{\url{https://github.com/cds-astro/ipyaladin}}, an interactive visualization tool based on Aladin Lite\footnote{\url{https://aladin.cds.unistra.fr/AladinLite/}} for individual assessment. 
By overlaying HSC-SSP optical images together with known redshifts from DESI\footnote{In the following, `DESI' refers to `DESI EDR SV3' unless otherwise specified.} or NED\footnote{NASA/IPAC Extragalactic Database: \url{https://ned.ipac.caltech.edu/}}, we confirmed faint but reliable detections associated with optical counterparts.

\begin{figure}[t]
    \centering
    \includegraphics[width=1\columnwidth]{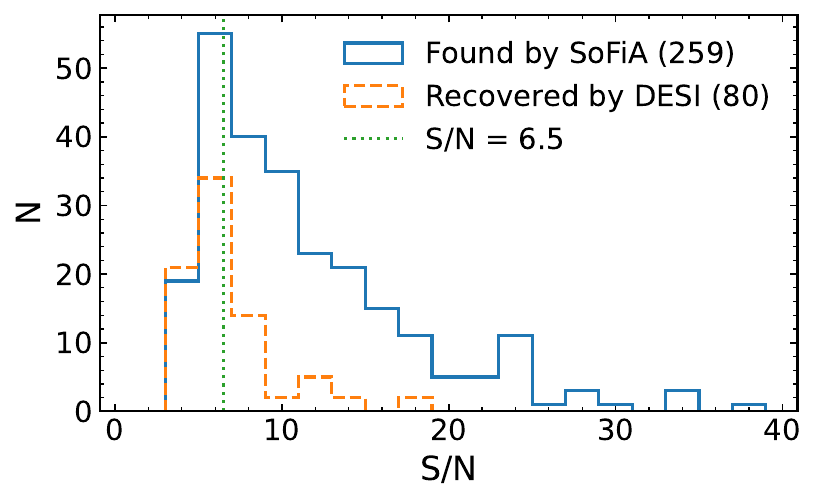}
    \caption{S/N distribution of the 339 \HI\ sources in the pilot \HD2\ catalog, shown for $3 \leq \mathrm{S/N} \leq 40$ for clarity. The blue solid histogram shows the sources identified by \SOFIA, and the orange dashed histogram shows the sources subsequently recovered through targeted searches at known DESI redshift positions. The green dotted vertical line indicates $\mathrm{S/N} = 6.5$, which separates high-S/N and low-S/N sources.
    }
    \label{fig:SNR_hist}
\end{figure}

The full identification process yielded 339 \HI\ sources, and all of them have at least one OC. This catalog includes two categories: (1) 259 sources identified and confirmed through the combined \SOFIA\ and human-vetting procedures, all with at least one optical counterpart; and (2) 80 additional sources initially missed by \SOFIA's linking algorithm---typically due to low signal-to-noise ratio (S/N) or complex morphology---but later recovered through targeted searches at positions of known DESI redshifts.
The S/N distributions of these two samples are compared in Figure~\ref{fig:SNR_hist}, showing that the DESI-recovered sources are predominantly at low S/N, consistent with them being missed by \SOFIA\ due to faint signals.
We note that a few sources with relatively high S/N were missed by \SOFIA. This is likely due to their locations near the edge of the observed field or the parameter settings in \SOFIA's linker process, which is acceptable for the pilot survey. For the ongoing full \HD2\ survey, we will release a more comprehensive blind \HI\ catalog with higher completeness.
To provide a clearer understanding of the data quality and reliability of these detections, we present some representative \HI\ spectra and optical images spanning a range of S/N values in Appendix~\ref{sec:optspec}.
All sources with $\mathrm{S/N} < 3$ have been discarded from the final catalog.

\subsection{Parameter Measurement}

Although \SOFIA\ is an effective automated tool for source extraction, it struggles to separate sources that are spectrally and spatially blended within FAST's large $3'$ beam and $1'$ grid. In such cases, the flux of neighboring sources becomes confused when \SOFIA\ applies multi-scale smoothing.
To address this, we apply an independent classification and parameter measurement procedure to derive the physical properties of the \HI\ sources. 
For sources that are partly spatially-blended but still visually separable (e.g., showing distinct peaks), we use the \texttt{segmentation} module in the Python package \texttt{photutils}\footnote{\url{https://photutils.readthedocs.io/en/stable/index.html}} to divide the zeroth-moment map and assign individual masks. 
For spectrally blended sources, extraction remains more difficult, so we classify these cases as confused and discuss their percentage in Section~\ref{sec:confuse}.
In addition, \HI\ sources that are split into multiple components by \SOFIA\ are identified and merged during the parameter measurement stage. The flux is measured over the entire profile, and duplicate entries are removed.

\subsubsection{Source Aperture and Flux} \label{sec:aperture}

The source aperture is the area encompassing signals, used to integrate the total flux of the source. 
In \SOFIA, this is represented by a three-dimensional mask around the detected signal. However, due to the large beam and pixel size of single-dish data, the mask often includes excess noisy pixels, reducing the S/N.
To address this, we estimate apertures using the curve-of-growth (CoG) method, widely applied in optical photometry. We employ \texttt{petrofit}\footnote{\url{https://petrofit.readthedocs.io/en/latest/index.html}} \citep{Geda2022} to construct the CoG and define the radius that encloses the total flux for most high-S/N sources, measured on a $25' \times 25'$ zeroth-moment ``postage stamp" map. 
For low-S/N sources, where the CoG is less reliable, we instead determine the aperture by fitting elliptical isophotes according to the flux contour.

The derived source aperture is larger than the FAST beam, ensuring that the full source flux is enclosed. Following \citet{Shostak1980}, we integrate the pixels within the aperture to measure the flux density ($S_\nu$ in Jy):
\begin{equation}
S_\nu = \frac{1}{\Omega_{A}} \sum_{A} s_\nu(x, y), 
\label{equ:Sv}
\end{equation}
where $s_\nu(x, y)$ is the flux density of each pixel (in \Jybeam) and $A$ is the aperture area. The summed flux density is normalized by the effective solid angle of the beam, $\Omega_A$, similar to the beam term in Eq.~3 of \citet{Haynes2018}.
For an ideal Gaussian beam with full width at half maximum (FWHM) $\theta_a$ and $\theta_b$, the full solid angle is
$\Omega_{\mathrm{all}} = (\pi \theta_a \theta_b) / (4 \ln 2)$. Since only discrete aperture pixels are summed, we instead use the solid angle contained within the aperture, $\Omega_A$. To compute $\Omega_A$, we evaluate the celestial separation of each pixel from the aperture center and apply a 1D Gaussian profile. For FAST's circular beam, we adopt a fixed FWHM of $3'$ to simplify the calculation\footnote{Most low-redshift sources lie in 1340 -- 1380 MHz, so a constant $3'$ beam size is assumed, without accounting for variations among the 19 beams.}.

After these steps, a low-order polynomial baseline is fitted to the integrated spectrum. The final flux density profile is obtained by integrating over the velocity channels containing the signal. Comparison with ALFALFA results in \citet{Jing2024} confirms the accuracy and consistency of our flux measurements.

\subsubsection{Velocity Width}\label{sec:width}

We mainly adopt two definitions of velocity width: $W_{50}$, following \citet[][hereafter \citetalias{Springob2005}]{Springob2005}, and $V_{85}$, based on the spectral centroid method of \citet[][hereafter \citetalias{Yu2020}]{Yu2020}.
The \citetalias{Springob2005} approach, used in ALFALFA \citep{Haynes2011}, measures the velocity range where the flux density falls to 50\% of the peak value (or the single peak, if only one is present), defining the line width $W_{50}$. 
We also provide $W_{20}$, the velocity range where the flux density falls to 20\% of the peak value in the catalog following the same method.
For low-S/N cases where this method is less robust, we adopt the automatic technique proposed by \citetalias{Yu2020} and further developed in \citet{Yu2022}. This method is fully automated and computationally more efficient than fitting-based approaches \citep{Ball2023}. Similar to the spatial CoG discussed in Section~\ref{sec:aperture}, the spectral CoG accumulates flux density symmetrically from both sides of the profile. The line width enclosing 85\% of the total flux is taken as $V_{85}$. 
When the $W_{50}$ auto-measurement  fails (mostly for sources with S/N $<6.5$), an interpolated value from $V_{85}$ could be implemented in the profile measurement.

\subsection{Optical Counterparts} \label{sec:optcp}

Having obtained the \HI\ source catalog, we now need to identify optical counterparts for each source. The observation region has an abundant coverage of extragalactic spectra provided by the SDSS, HectoMAP, and DESI EDR SV3. 
DESI is capable of identifying fainter galaxies with newly determined redshifts that surround the host galaxy, thereby uncovering previously overlooked confusion or incorrect matching instances.
In the following, we present the method for matching OCs and quantify the rate of correct matches in identifying these counterparts. Additionally, we evaluate the percentage of confusion among the matches.

\subsubsection{Methods}\label{sec:match}

The method to match the best OC of our \HI\ sources is modified from \citet{Zhang2024}. 
First, we search for the optical counterparts within a radius of $3'$ of the \HI\ source.
For spectroscopic redshift surveys, we apply a velocity restriction of $\pm 300$~\kms to include the galaxies that are confused with the \HI\ source. In contrast, for photometric redshift surveys, we utilize a larger range of $\pm 7000$~\kms, which corresponds to the $1\sigma$ scatter of the velocity difference between photometric and spectroscopic redshifts at low redshift.
When no optical object satisfies the matching criteria, it indicates that no optical objects correspond to the \HI\ sources. 
For those that do meet the criteria, the probability of each optical target is defined as the product of four parts: coordinates, redshifts, magnitudes, and color.
Then we multiply them to obtain the OC probability of the objects surrounding the \HI\ source,
\begin{equation}
    P_\mathrm{OC} = P_\mathrm{coord}P_\mathrm{vel}P_\mathrm{mag}P_\mathrm{color}.
\label{equ:Poc}
\end{equation}

Equation~\ref{equ:Poc} features four probability terms.
The probabilities from spatial offset and velocity separation both follow normal distributions.
For photometric data, the term $P_\mathrm{vel}$ is ignored, and the magnitude probability is defined by
$P_\mathrm{mag} =  \exp \left[ -\left( {D_i}/{3'} \right)^2 N(m<m_i) \right]$,
where $D_i$ is the separation between \HI\ source and the $i$th OC coordinates in arcmin, and $N(m<m_i)$ is the number of objects brighter than it in $3'$ field \citep{Downes1986, Zhang2022a, Zhang2024}.
For spectroscopic surveys, the magnitude probability is
$P_\mathrm{mag} = \exp \left[-\left( {D_i}/{3'} \right) ^2 \right] F(m_i)$, where $F(m_i)$ is the normalized flux depending on r-band magnitude ($m_i$) to provide an exponential weight, and here we use $F(m_i) \propto 10^{m_i/-2.5}$.
The last term, $P_\mathrm{color}$, is based on the color of the galaxies: blue galaxies usually contain more gas, so we prefer to give blue galaxies a larger weight based on their color.
Similar to $P_\mathrm{mag}$, it also follows an exponential relation, like $P_\mathrm{color} \propto 10^{(g-r)/-2.5}$.
Note that $P_\mathrm{color}$ is only applied to galaxies with $r > 17$ mag, since the nuclei or bulges of luminous galaxies tend to have redder colors that may bias the weighting.

\begin{figure*}[ht!]
\plotthree{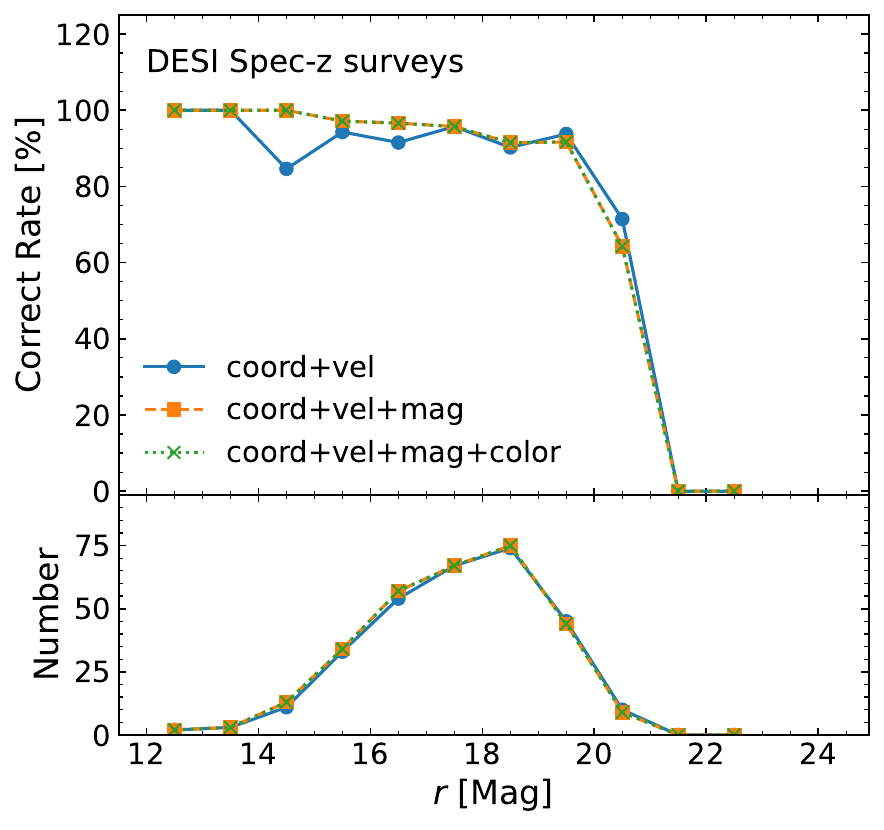}{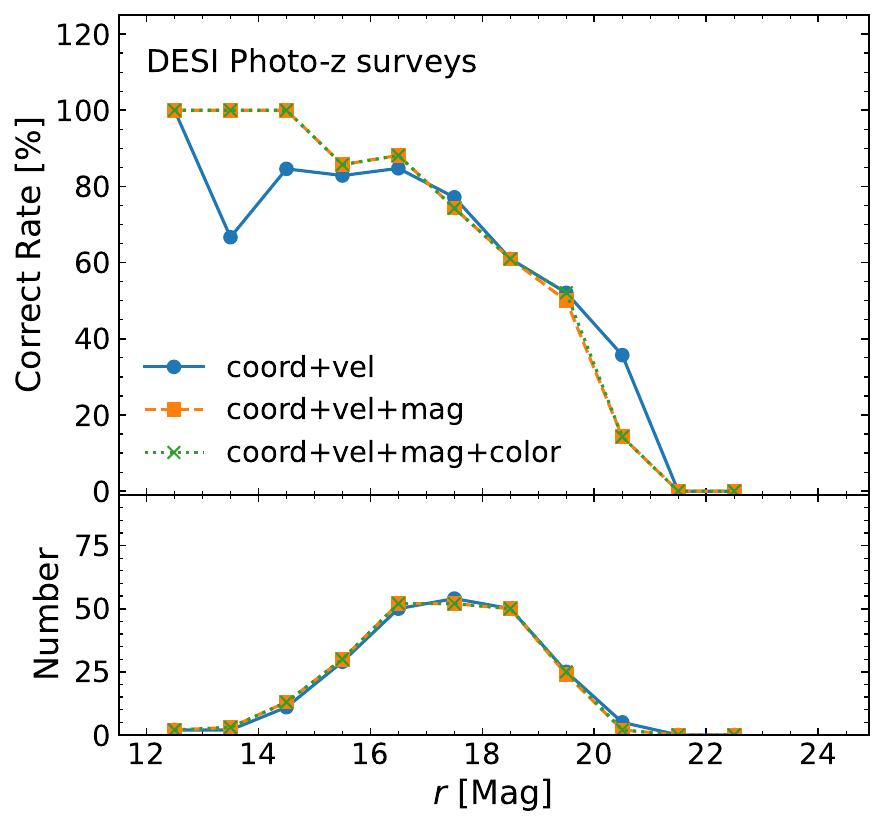}{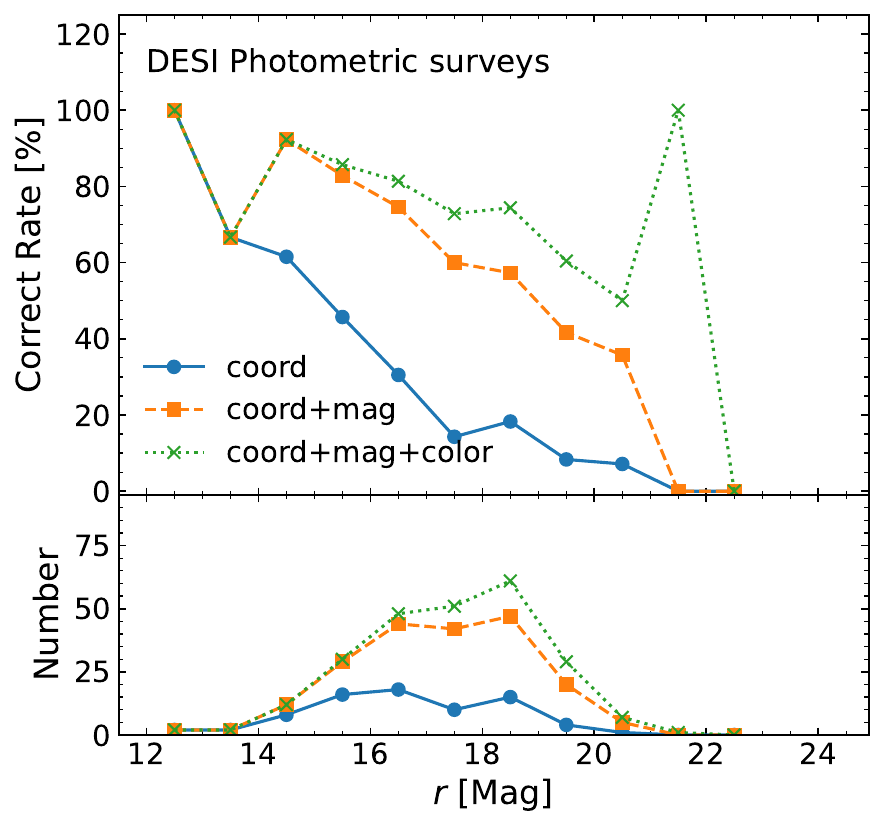}
\caption{The OC correct rate as a function of $r$-band magnitude for DESI spectroscopic (left), photometric redshift (middle), and pure photometric (right) surveys, comparing different combinations of probability terms. 
Each panel compares three combinations of probability terms: coordinates and velocity only, or coordinates only for the pure photometric case (blue circles, solid line); 
with the addition of $P_\mathrm{mag}$ (orange squares, dashed line); 
and further adding $P_\mathrm{color}$ (green cross, 
dotted line). 
The lower panels show the number of matched OCs in each magnitude bin.}
\label{fig:OC_term}
\end{figure*}

The four terms are approximately equally weighted: the coordinate and velocity terms are normalized to unity at the peak of their respective Gaussian distributions; the magnitude term is normalized such that the brightest object in the field has $P_\mathrm{mag} = 1$; and the color term is also normalized to the value of the bluest OC candidate.
To evaluate the contribution of each term, we perform a systematic test using DESI spectroscopic, photometric redshift, and pure photometric surveys separately, comparing the OC correct rate when progressively adding terms: 
(1) coordinates (and velocity) only, (2) adding $P_\mathrm{mag}$, and (3) further adding $P_\mathrm{color}$. 
The correct rate is defined as $N_\mathrm{correct} / N_\mathrm{total}$, where $N_\mathrm{correct}$ is the number of \HI\ sources that are correctly matched with the best OC according to known redshifts, supported by human assessments based on images from HSC, and $N_\mathrm{total}$ is the total number of \HI\ sources within the DESI sky coverage. 
The results are shown in Figure~\ref{fig:OC_term}.
For spectroscopic and photometric redshift surveys, the combination of $P_\mathrm{coord}$, $P_\mathrm{vel}$, and $P_\mathrm{mag}$ is already sufficient to achieve a high correct rate, and the addition of $P_\mathrm{color}$ does not significantly affect the results, since the other terms already strongly constrain the OC selection. 
Nevertheless, we retain the color term as it provides a meaningful improvement for pure photometric surveys where no redshift information is available. 

Finally, the optimal OC is defined as the optical object with the maximum total OC probability.

\subsubsection{OC Matching Rate and Correct Rate}\label{sec:rate}

\begin{figure*}[ht!]
\plotthree{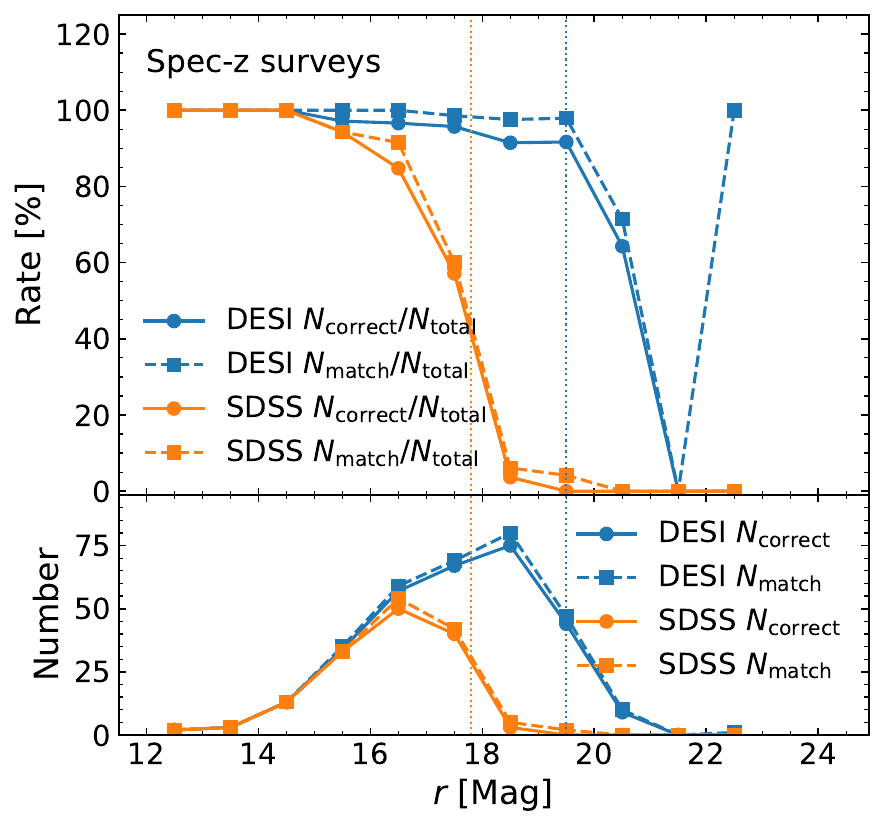}{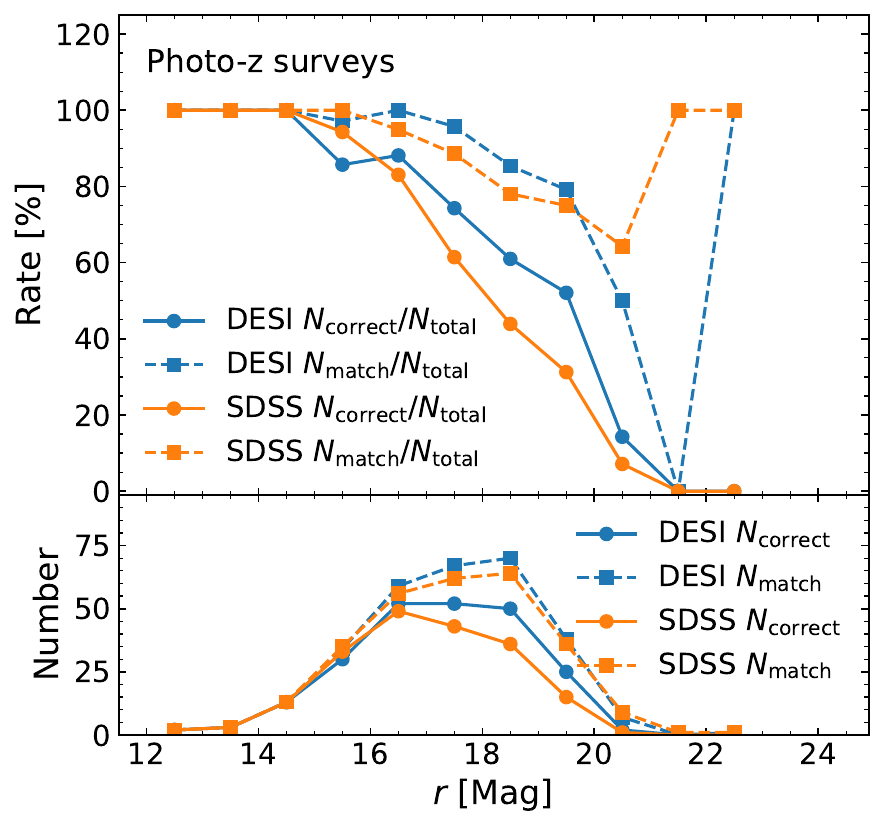}{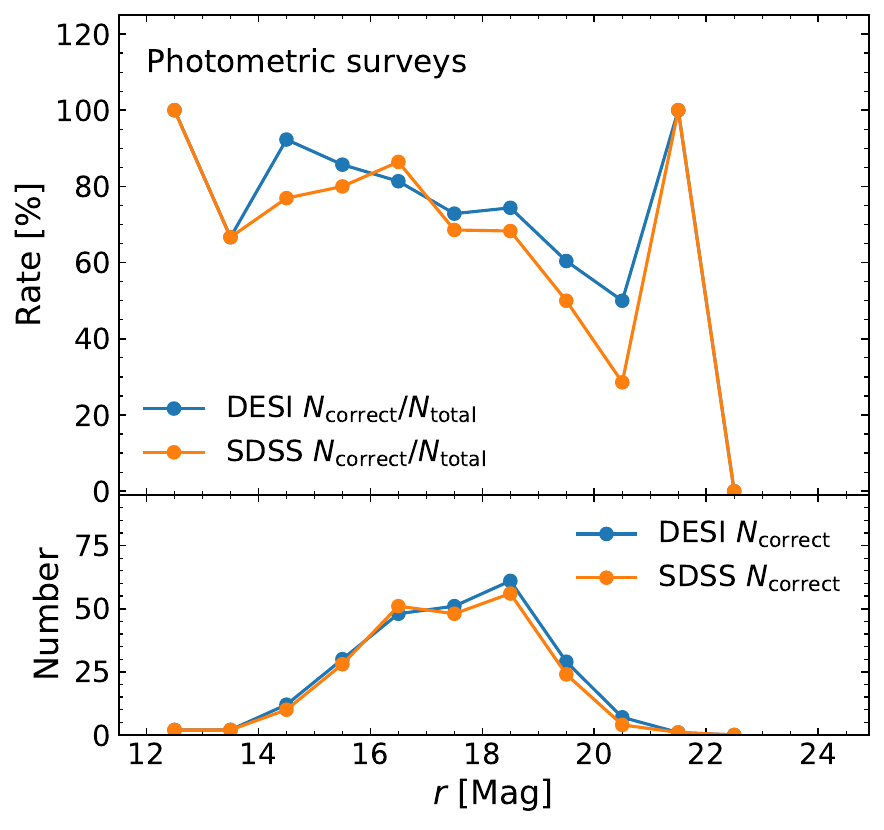}
\caption{The matching rate and correct rate of optical counterparts using DESI EDR/SDSS spectroscopic redshift surveys (left), photometric redshift surveys (middle) or photometric surveys (right). 
The $x$-axis presents the $r$-band visual AB magnitude of the correct optical counterpart.
The dashed curves indicate the matching rate, and the colored solid lines correspond to the correct rate. 
The lower panels illustrate the matching or correct numbers of each magnitude bin.
The vertical dashed lines illustrate the magnitude at which the spectroscopic objects become incomplete.
The matching rate and correct rate of spectroscopic redshift surveys are higher than those of photometric redshift and pure photometry at $r <$ 20 mag. Furthermore, it is evident that the correct rate of SDSS spectroscopy declines earlier than that of DESI due to its shallower depth.
}
\label{fig:rate}
\end{figure*}

With the probabilities of optical counterparts for the \HI\ source established, we can assess the accuracy of the most probable optical counterpart using various optical surveys, such as SDSS and DESI.
Figure~\ref{fig:rate} presents the matching rate and correct rate derived from DESI and SDSS spectroscopic redshift (spec-$z$) surveys, photometric redshift (photo-$z$) surveys and pure photometric surveys. 
Similar to the correct rate, we also define the matching rate as $N_\mathrm{match} / N_\mathrm{total}$, where $N_\mathrm{match}$ represents the number of \HI\ sources matched with OCs.

For spec-$z$ surveys, the matching criteria are $3'$ and $\pm 300 \, \mathrm{km~s^{-1}}$.
For bright and predominantly large galaxies, matching using spec-$z$ maintains a high accuracy of over $90\%$ until the redshift completeness becomes inadequate. The $r$-band spectroscopic completeness turning points are 17.8 mag for SDSS \citep{Strauss2002} and 19.5 mag for the DESI Bright Galaxy Survey (BGS) bright sample \citep{Hahn2023}, as indicated by the vertical lines in the left panel of Figure~\ref{fig:rate}. As expected, DESI exhibits greater photometric depth and encompasses more spectroscopic objects than SDSS, resulting in significantly better matching and correct rates for galaxies with $r$-band magnitudes between 17 and 19.5 mag. For galaxies fainter than 20.5 mag, both SDSS and DESI struggle to match any OCs accurately.

For photo-$z$ surveys, the matching criteria are relaxed to $3'$, $\pm 7000 \, \mathrm{km~s^{-1}}$. Similarly to spec-$z$ surveys, we present the matching rate and correct rate in the middle panel of Figure~\ref{fig:rate}. 
The DESI photo-$z$ exhibits lower matching and correct rates compared to spec-$z$ for galaxies with $r < 20.5$ mag due to the large uncertainties associated with photometric redshifts. 
In contrast, the SDSS can provide a larger number of objects matched with a correctness of approximately 40\% to 80\% for magnitudes between 17 and 19 mag, although this is still lower than the matching rate of DESI spec-$z$.

For photometric surveys, we only plot $N_\mathrm{correct} / N_\mathrm{total}$ in the right panel of Figure~\ref{fig:rate} since every \HI\ source would match at least one OC. 
As apparent magnitude increases, the correct rate gradually declines from 80\% to 40\%, with DESI maintaining a slightly higher correct rate than SDSS.
When galaxies have magnitudes fainter than 20 mag, the limited samples lead to significant fluctuations in the final two data points, causing the matching results to become unreliable.

\subsubsection{Confusion Sample Percentage} \label{sec:confuse}

When matching the best OC, we often encounter instances where the \HI\ source has multiple OCs. 
Due to DESI's greater depth and higher completeness at the faint end of the magnitude, confusion is likely to be a common occurrence for FAST's single-dish beam of $\sim$3 arcmin \citep{Jones2015, Jones2016}.
For most unresolved cases, the confused mass contributed by small neighboring galaxies could be ignored, as they may not have much \HI\ gas.
However, galaxies with comparable optical sizes may suffer from an evident bias. The large FAST beam likely measures the total \HI\ emission of both galaxies.
Confusion will cause the wrong estimation of the galaxies' \HI\ mass, so we investigate the percentage of confusion cases in our \HI\ samples.

\begin{figure}[ht!]
\plotone{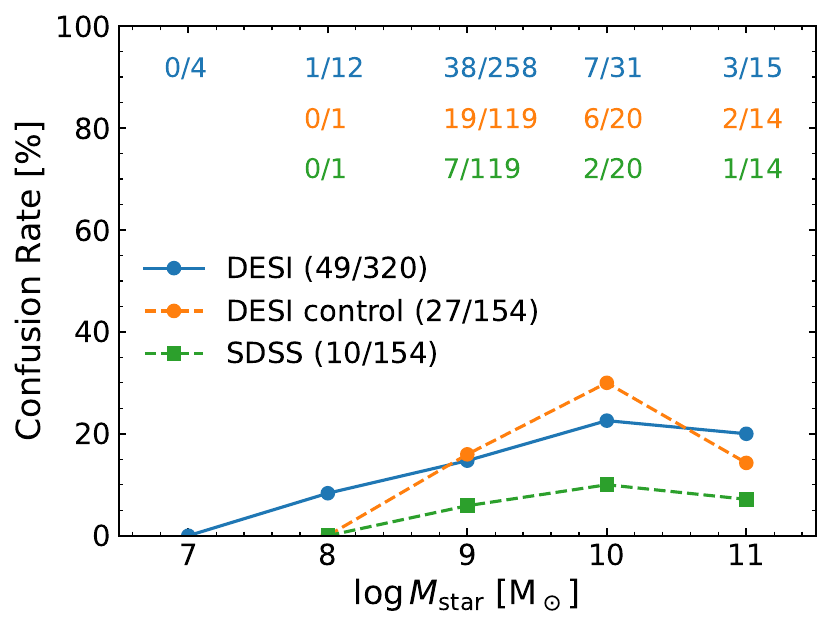}
\caption{The confusion rate in the pilot \HD2\ sample is analyzed by binning the stellar mass of the best-observed counterparts associated with each \HI\ source. There are 320 \HI\ sources within the DESI's coverage, of which 49 exhibit confusion (DESI, blue solid line with circle markers). Among the 154 galaxies that contain OCs from SDSS, ten are identified as confused (SDSS, green dashed line with square markers). In comparison, the control sample utilizes DESI, which includes 27 confused sources (DESI control, orange dashed line with circle markers). The colored numbers above indicate the count of confused sources and total sources in each bin.
}
\label{fig:confuse_rate}
\end{figure}

To describe its influence, we evaluate the confusion via the number, $N(P_\mathrm{OC} > \alpha P_\mathrm{OC~best}) > 1$, where $P_\mathrm{OC}$ is the probability of optical targets that meet the matching criteria.
We chose $\alpha=0.1$ to define the most significant confusion situation, which indicates the number of OCs with a probability larger than 10\% of the best OC. 
We refer to sources satisfying this criterion as ``severely confused'', as these are the cases most likely to result in an overestimate of the \HI\ mass of the primary galaxy.
When analyzing isolated galaxies, we will exclude those severely confused sources.
Figure~\ref{fig:confuse_rate} quantifies the percentage of severely confused samples in the pilot \HD2\ catalog, categorized by the stellar mass of the best optical counterparts associated with each \HI\ source. We compare the confusion rates for different optical spectroscopic surveys. 
For DESI, the confusion rate is nearly 20\% for galaxies with stellar masses of approximately $M_\star \sim 10^{8.5-10}$ \Msun. When we consider the overlapped sample from DESI and SDSS as a control, DESI reveals a higher confusion rate than SDSS. 
The confusion rate using SDSS is underestimated at approximately 10\%, indicating that SDSS has missed the spectra of faint galaxies that likely contain \HI.

To evaluate whether our probability-based confusion definition exaggerates the practical effects on \HI\ mass measurements, we also estimate the \HI\ mass of each OC candidate using the \HI\ gas fraction scaling relation based on NUV$-r$ \citep{Saintonge2022}, applying a threshold of 30\% of the \HI\ mass of the primary OC as an alternative confusion criterion. Comparing the two definitions for sources with available NUV$-r$ colors, we observe consistent overall confusion rates of $\sim$12\%--15\%. 
Only 5\% and 8\% of sources exhibit discrepant classifications, being labelled as confused only by the \HI\ scaling relation or by the optical method, respectively. 
Given the scatter in \HI\ scaling relations, we find that our probability-based definition does not substantially overstate the impact of confusion on the derived \HI\ masses.

\subsection{\HI\ Stacking} \label{sec:stack}

\subsubsection{Galaxy Spectral Stacking}
Our stacking parent sample is drawn from the stellar mass and emission line measurements for galaxies in DESI EDR\footnote{\url{https://data.desi.lbl.gov/doc/releases/edr/vac/stellar-mass-emline/}} \citep{Zou2022, Zou2024}, which contains $\sim$1400 galaxies with reliable redshifts in the 10-\deg2 field.
First, we select the highly complete BGS bright and faint samples, which include approximately 1000 galaxies.
To minimize the impact of confusion, we remove galaxy pairs with an angular separation $r<3'$ and a velocity deviation of $|\Delta V|<300$~\kms, and also galaxies overlapped with other galaxies with \HI\ detections.
The remaining 425 isolated galaxies without \HI\ detections or companions are visually inspected, and those with poor baselines or residual RFI in their stacked spectra are excluded from the analysis.
We then restrict the redshift range to $z<0.09$, leaving 376 galaxies. Further cuts are applied in stellar mass, stellar surface density, and specific star formation rate (sSFR) to define subsamples with sufficient binning statistics:
\begin{align}
& 8.1<\log M_{\star}~[\mathrm{M}_{\odot}]<11.2, \\
& 6.3<\log \mu_{\star}~[\mathrm{M}_{\odot}~\mathrm{kpc}^{-2}]<9.2, \\
& -13.1<\log \mathrm{sSFR}~[\mathrm{yr}^{-1}]<-8.7.
\end{align}
The final sample consists of 354 galaxies: 229 direct \HI\ detections and 125 stacking sources without significant confusion, which are used to study the scaling relations.

The stacking procedure follows the approach of \citet{Fabello2011} and \citet{Healy2019}.
For each galaxy, we extract a spectrum from the \HI\ data cube using its optical coordinates and velocity. For galaxies without direct \HI\ detections, we adopt a fixed circular aperture of $6'$ diameter, corresponding to twice FAST's beam size.
We then cut a velocity range of 4000~\kms\ centered on the DESI optical redshift and measure the flux density following the methods in Section~\ref{sec:aperture}.
The spectra are grouped by galaxy properties (e.g., stellar mass, color). Each spectrum is shifted to the rest frame and aligned at zero velocity. 
Because targets within a bin may span a wide range of redshift or stellar mass, we stack their \HI\ mass (or \HI\ gas fraction) directly instead of the \HI\ flux, weighting by $1 / \sigma_\mathrm{rms}^2$.
From the stacked spectra, we derive the average $M_\mathrm{HI}$ or \HI\ fraction using the same spectral CoG method as applied to direct detections. The error is estimated by employing bootstrap resampling on 80\% of the galaxies within the bin \citep{Fabello2011}.

\subsubsection{Group Stacking} \label{sec:group_stack}

As FAST's large beam introduces non-negligible confusion in stacking \citep{Jones2016}, a straightforward approach is to co-add the \HI\ spectra of entire groups rather than individual galaxies.
However, FAST's resolution makes it difficult to distinguish centrals from satellites, so we stack the total group \HI. 
The group stacking is based on the DESI Legacy Survey DR9 group catalog\footnote{\url{https://gax.sjtu.edu.cn/data/DESI.html}} \citep{Yang2021}, updated with DESI EDR's 1\% spectroscopic redshifts. Details of the group finder algorithm are given in \citet{Yang2005, Yang2007}.

Following the same procedures as for individual galaxies, we select groups without significant overlap. For massive halos within our limited volume, we retain groups with only minor overlaps.
After removing groups affected by poor baselines or residual RFI, a total of 680 groups remain for stacking.
We extract spectra within circular apertures centered on the group positions provided by \citet{Yang2021}.
The aperture is defined by the angular size of the group, which is based on twice the halo radius, $r_{180}$. For halos with an angular diameter of less than $6'$ and without direct \HI\ detections, we instead adopt an aperture of $6'$ to encompass the flux within the group \citep{Guo2020, Guo2021}.
The subsequent steps follow the same procedures as those for individual galaxy stacking. From the final stacked spectra of groups, we then measure the average \HI\ mass.

\section{Results}\label{sec:result}

We detect 339 \HI\ emission lines at $z<0.09$ based on the \SOFIA\ 2 and optical target searching.
Table~\ref{tab:catalog} presents the measured parameters in the low-redshift data release of the pilot survey. 
We only select sources with velocity $>$ 300 \kms, ensuring that they are predominantly extragalactic \HI\ signals, not galactic high-velocity clouds, OH megamasers, or radio recombination lines after the human examination.
The contents of Table~\ref{tab:catalog} are introduced in Appendix~\ref{sec:catalog}.

\subsection{Comparison with other Surveys}

Table~\ref{tab:compare} shows that our 10-\deg2 \HD2\ pilot survey reaches an average source density of 33.9 per deg$^2$, about six times higher than previous wide \HI\ surveys.

We compare our results with existing \HI\ surveys in the same sky area. FASHI reported 39 sources with S/N $>$ 5, and we detect 298 sources with S/N $>$ 5 in the same area. 
We recover 37 of them; for the remaining two, one is separated into two sources in our catalog, and one is not detected in our data cube.
In this pilot test, the deeper observation and the three-pass scan strategy with DESI OC identifications enable us to detect nearly an order of magnitude more \HI\ sources than FASHI.

\begin{deluxetable*}{lccccccccc}
\tabletypesize{\scriptsize}
\tablewidth{0pt} 
\tablecaption{Comparison of some \HI\ surveys. 
Here we list another four surveys with their beam size, sky area, maximum redshift, rms at 20~\kms\ resolution, 3$\sigma$ column density limitation ($\log N_\mathrm{HI}^\mathrm{lim}$), integration time per beam solid angle ($t_\mathrm{int}$), source detection number ($N_\mathrm{det}$), detection number density ($n_\mathrm{det}$), and references.}
\label{tab:compare}
\tablehead{
\colhead{Survey} & \colhead{Beam size} & \colhead{Area} & \colhead{$z_\mathrm{max}$} & \colhead{rms$^a$} & \colhead{$\log N_\mathrm{HI}^\mathrm{lim}$} & \colhead{$t_\mathrm{int}$} & \colhead{$N_\mathrm{det}$$^b$} & \colhead{$n_\mathrm{det}$} & \colhead{Reference} \\
\\
\colhead{} & \colhead{(arcmin)} & \colhead{($\mathrm{deg^2}$)} & \colhead{} & \colhead{($\mathrm{mJy~beam^{-1}}$)} & \colhead{($\log \mathrm{cm}^{-2}$)} & \colhead{($\mathrm{sec~beam^{-1}}$)} & \colhead{} & \colhead{($\mathrm{deg^{-2}}$)} \\
} 

\colnumbers

\startdata 
HIPASS & 15 & 30000 & 0.04 & 12.3 & 18.0 & 450 & 5500 & 0.2 & $^c$ \\
ALFALFA & 3.5 & 6920 & 0.06 & 1.6 & 18.4 & 48 & 31500 & 4.6 & $^d$ \\
FASHI DR1 & 2.9 & 7600 & 0.09 & 0.4 & 18.0 & 50$^e$ & 41700 & 5.5 & {\citet{Zhang2024}} \\
\HD2\ pilotDR & 2.9 & 10 & 0.09 & 0.2 & 17.7 & 440 & 339 & 33.9 & This paper \\
WALLABY PDR1 & 0.5 & 180 & 0.08 & 0.7 & 19.7 & 57600$^f$ & 592 & 3.3 & {\citet{Westmeier2022}} \\
\enddata

\tablecomments{
$^a$ Based on map noise or detection limit, not spectral rms. \\
$^b$ We note that different surveys apply various S/N cuts. For instance, both ALFALFA and our survey require S/N almost $>3$, FASHI imposes a limit of S/N $>5$, and WALLABY restricts to S/N $>4.5$. \\
$^c$ \citet{Barnes2001, Meyer2004, Wong2006}. \\
$^d$ \citet{Giovanelli2005, Giovanelli2016, Haynes2018}. \\
$^e$ The integration time of FASHI is not uniform because some areas have been scanned multiple times. In the drift mode, integration time also depends on declination. \\
$^f$ Integration time of one pointing. 
}
\end{deluxetable*}

\begin{figure}[ht!]
\plotone{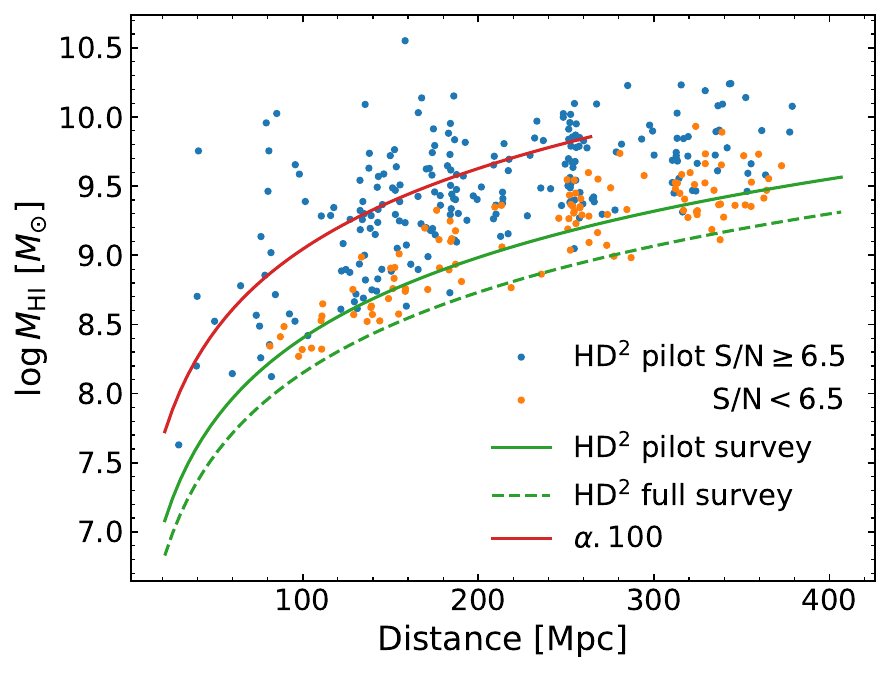}
\caption{Spaenhauer diagram for the low-redshift samples ($z<0.09$). The blue dots represent 230 \HI\ sources with a higher S/N of $\geq 6.5$, while the orange dots indicate 109 sources with a lower S/N. 
The red curve represents the $5\sigma$ mass limit of the ALFALFA samples ($\alpha. 100$), while the green solid and dashed curves correspond to the \HD2\ pilot survey and the full \HD2\ survey in prediction, respectively.
}
\label{fig:Mhi-D}
\end{figure}

Figure~\ref{fig:Mhi-D} shows the \HI\ mass-distance distribution of our pilot survey. Sources are plotted as colored dots, with 230 high-S/N detections in blue and 109 low-S/N detections in orange. The red curve marks the $5\sigma$ mass limit of ALFALFA, while the green curve corresponds to our pilot survey. The flux limit is calculated following \citet{Ponomareva2023},
\begin{equation}
S_{\lim}=5\sigma_{\mathrm{rms}}~\delta v \sqrt{\frac{W_{50}}{\delta v}},
\label{equ:flux_lim}
\end{equation}
where we adopt a representative line width of $W_{50}=200$~\kms\ and use the median rms of each survey (ALFALFA: 2.2 mJy; \HD2\ pilot survey: 0.5 mJy; full \HD2\ survey: 0.3 mJy) at a resolution of $\delta v\sim10$ \kms. Substituting $S_{\lim}$ into Equation~\ref{equ:himass} yields the \HI\ mass limit.
Compared with ALFALFA, our pilot survey achieves higher sensitivity and extends to higher redshifts but over a smaller sky area. 
 
It also enables the detection of galaxies with lower \HI\ masses at the same distance.
We note that the mass limitation curves assume a fixed linewidth and rms; thus, some high-S/N sources fall below the curves if their profiles are narrower or their local noise is lower. The low-S/N sources scatter around the green curve, consistent with expectations.
For the full \HD2\ survey, the spectral rms will decrease to 0.3 mJy at 10 \kms, which is approximately 1.6 times deeper than the pilot survey.

\subsection{\HI\ Source Detection Rate} \label{sec:det_rate}

\begin{figure*}[ht!]
\plotone{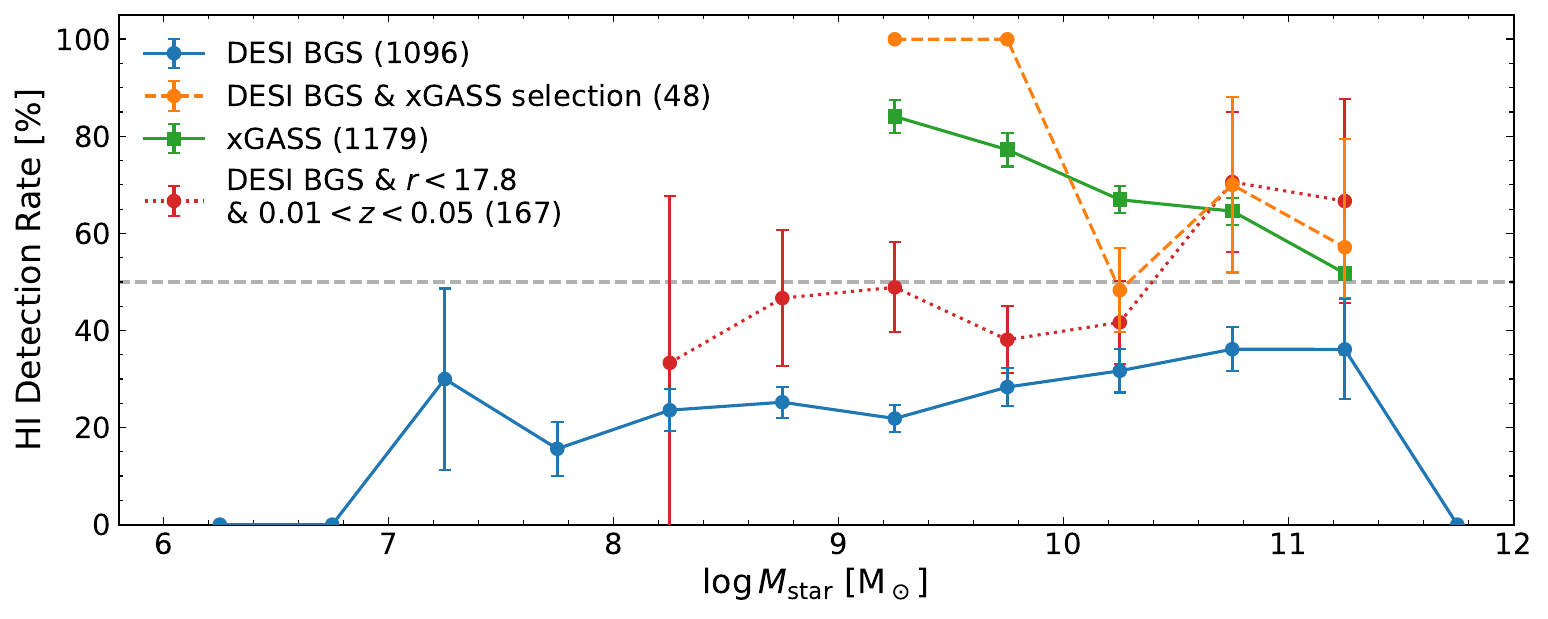}
\caption{The \HI\ detection rates of the DESI BGS sample are analyzed by binning the data according to stellar mass. Within the DESI BGS sample, the detection rate is approximately 30\% (the blue solid line with circular markers, 1096 galaxies). 
When applying the same selection criteria as used in xGASS, the \HI\ detection rate increases (the orange dashed line with circular markers, 48 galaxies), approaching that of xGASS (the green line with square markers, 1179 samples) at the high-mass end. 
If we restrict the redshift range to $0.01 < z < 0.05$, while maintaining the SDSS magnitude limit of $r < 17.8$ mag for the DESI BGS samples (red dotted line with circular markers, 167 galaxies), the \HI\ detection rate reaches nearly 50\% at $\log M_\star/\mathrm{M_\odot} < 10.5$, and is comparable to that of xGASS for massive galaxies.
The error bars are estimated using a bootstrapping method, and the dashed gray line indicates the 50\% detection rate. 
}
\label{fig:hidet_rate}
\end{figure*}

One of the key strengths of the final \HD2\ survey will be the high \HI\ source detection rate based on DESI data, and the pilot survey of \HD2\ offers an opportunity for estimation. We present the \HI\ source detection rate in Figure~\ref{fig:hidet_rate}, where it shows a nearly uniform detection rate of approximately 30\% for the DESI BGS galaxies, including both bright and faint samples. 

We compare our findings with the xGASS sample, which is the combination of GASS and GASS-low defined by the following criteria \citep{Catinella2010, Catinella2018}:
\begin{enumerate}
    \item Stellar mass range: $9.0 < \log M_\star/\mathrm{M_\odot} < 10.2$ for GASS-low; $10.0 < \log M_\star/\mathrm{M_\odot} < 11.5$ for GASS;
    \item Redshift range: $0.01 < z < 0.02$ for GASS-low; $0.025 < z < 0.05$ for GASS;
    \item $r$-band apparent magnitude from SDSS: $r < 17.8$ mag;
    \item UV photometry: FUV and NUV with magnitudes less than 21 mag or 23 mag, respectively.
\end{enumerate}
When applying the same selection criteria to the DESI BGS sample, but without the gas mass fraction limit, we expect the \HI\ detection rate to be higher, approaching the results seen in the xGASS survey. 
However, due to limitations in the survey volume, only 48 galaxies meet these criteria, resulting in a fluctuating detection rate above 50\%. Consequently, only two galaxies within the range of $9 < \log M_\star/\mathrm{M_\odot} < 10$ and $0.01 < z < 0.02$ have no error bars plotted in the orange dashed line.

The selection criteria for xGASS are strict, whereas the full BGS sample selection is excessively loose. 
To demonstrate the potential of our \HD2\ survey, we broaden the redshift range to $0.01 < z < 0.05$, while still adhering to the SDSS's magnitude limit of $r < 17.8$ mag for the DESI BGS sample. 
We find a detection rate of approximately 70\% for massive galaxies, which is similar to the detection rates observed in xGASS.
For those with $\log M_\star/\mathrm{M_\odot} < 10.5$, the \HI\ detection rate can reach nearly 50\%. 
But the full \HD2\ survey will be 1.6 times deeper than the pilot survey, and with a larger spectroscopic coverage from DESI, we will then be able to construct a complete optical sample selected by stellar mass and redshift, subsequently extending this sample to the low-mass end.

\subsection{\HI\ Detection Distribution} \label{sec:det_distribution}

\begin{figure*}[ht!]
\plottwo{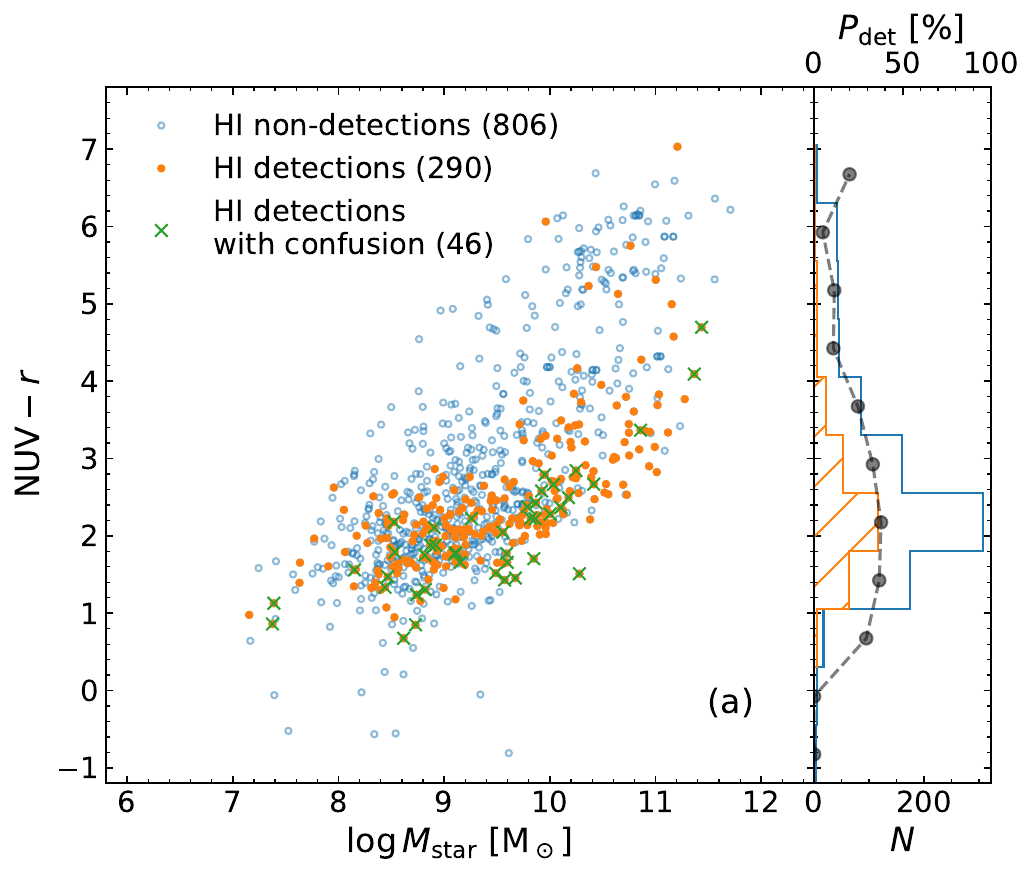}{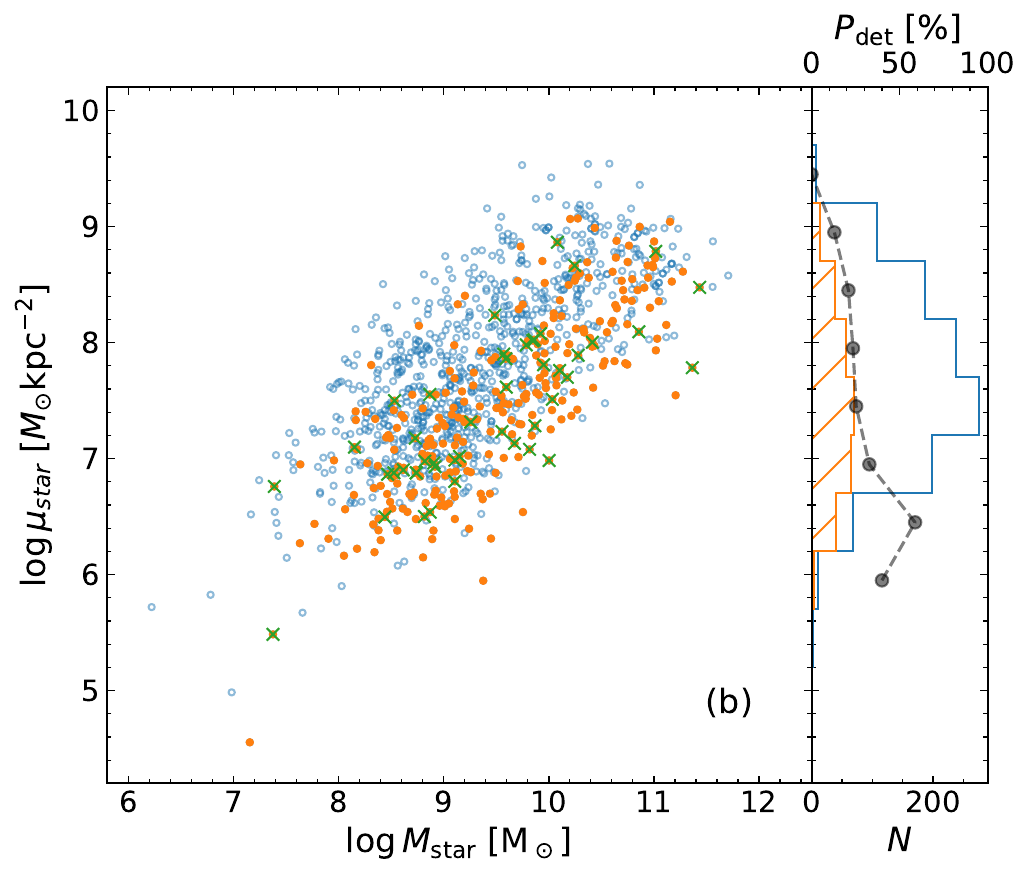}
\plottwo{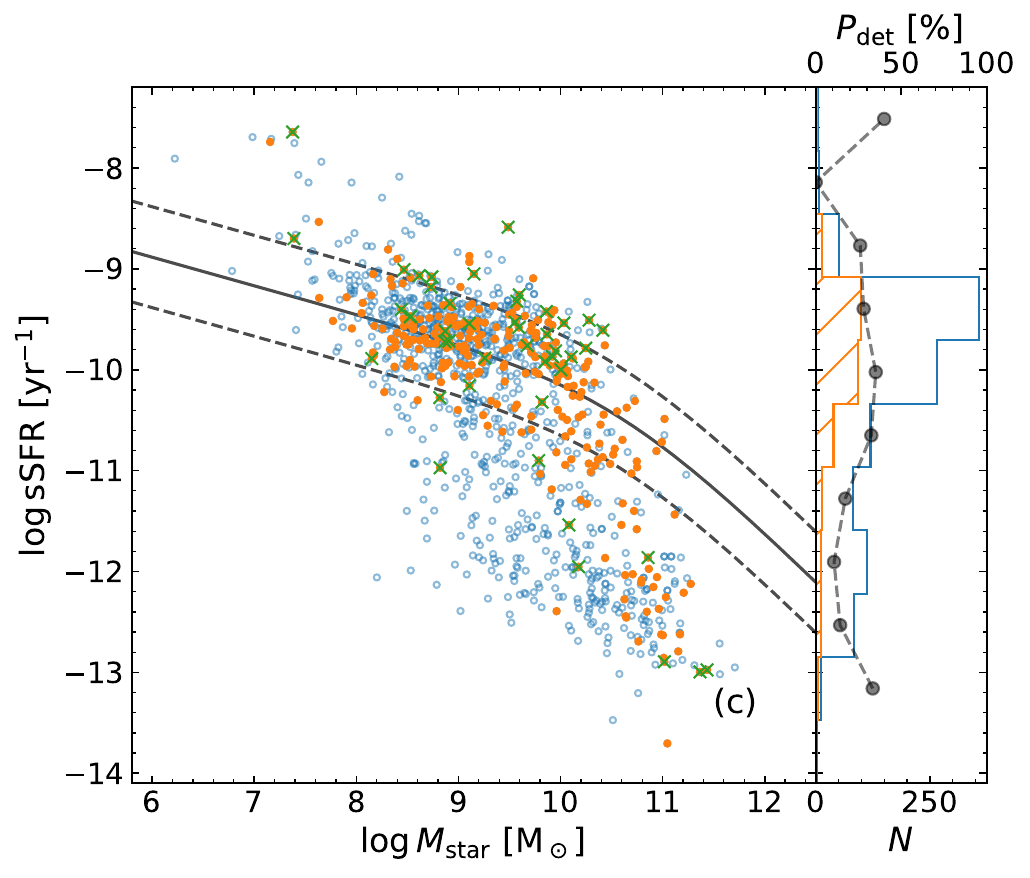}{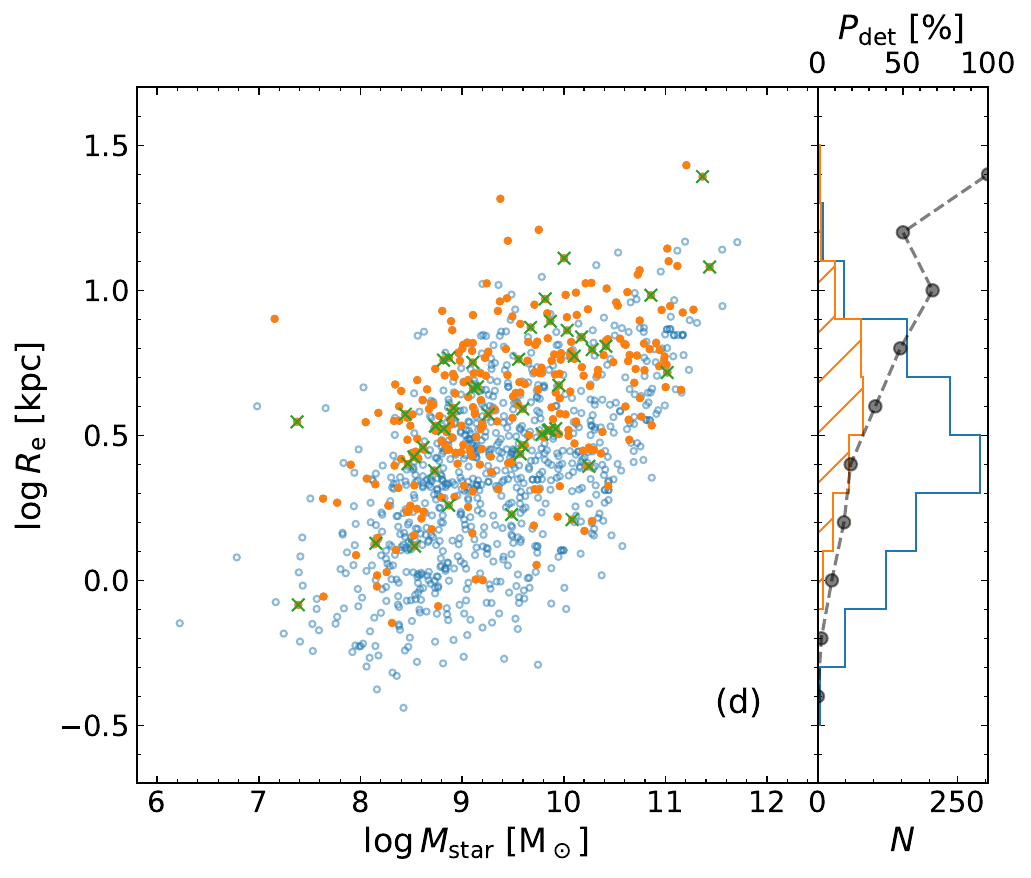}

\caption{The distribution of (a) $\mathrm{NUV}-r$, (b) stellar mass surface density, (c) specific star formation rate, and (d) half-light radius as a function of stellar mass for \HI\ detections and non-detections in the DESI BGS sample is shown. The orange dots represent galaxies with \HI\ detections, the green crosses indicate \HI\ detections flagged as confusion cases, while the blue open dots denote those without \HI. 
In panel (c), the black solid line shows the star-forming main sequence (SFMS), while the black dashed lines mark the $\pm 0.5$\,dex range around the SFMS.
The vertical histograms display the total number of DESI BGS galaxies (blue steps) and the number of galaxies with confirmed \HI\ detections (orange shadow) within each bin along the $y$-axis. The gray dots connected by dashed lines indicate the percentage of \HI\ detections in the entire DESI BGS sample from our survey.}
\label{fig:cmd}
\end{figure*}

Before combining \HI\ data with optical information to conduct scaling relations, we first examine the distribution of \HI\ detections and non-detections in the optical parameter space.  
In Figure~\ref{fig:cmd}, we show (a) the $\mathrm{NUV}-r$, (b) the stellar mass surface density ($\mu_{\star}$), (c) sSFR, and (d) the half-light radius ($R_\mathrm{e}$) as functions of stellar mass.
Stellar properties, including $M_\star$ and SFR, are derived from SED fitting \citep{Zou2022, Zou2024}. 
The stellar mass surface density is defined as $\mu_{\star} = M_\star/(2 \pi R_\mathrm{e}^2)$ \citep{Brown2015}, where $R_\mathrm{e}$ is the half-light radius from the DESI DR9 tractor catalog \citep{Dey2019} measured using the combination of $g$, $r$, and $z$ bands, converted to kpc. 
The sSFR is given by SFR/$M_\star$, and the $\mathrm{NUV}-r$ is computed from galactic extinction-corrected \textit{GALEX} NUV \citep{Bianchi2014} and BASS $r$-band magnitudes \citep{Zou2017}. 
We divide the DESI BGS galaxies into \HI\ detections and non-detections after matching their optical counterparts. For sources with optical confusion, we mark only the best optical counterpart as \HI\ detected and overplot them as green crosses.

Among these optical properties, we find that the \HI-detected galaxies span nearly the entire parameter space of the DESI sample.
In panel (a) of Figure~\ref{fig:cmd}, we present the stellar mass versus the $\mathrm{NUV}-r$ diagram, showing that most DESI BGS galaxies, as well as those in our pilot survey, are biased toward the blue population. Blue galaxies have a higher \HI\ detection rate, reaching up to 40\%, as expected. But for the red side, we still have a few detections.  
In panel (b), galaxies with $\log \mu_{\star}/(\mathrm{M}_{\odot}~\mathrm{kpc}^{-2}) < 8.5$ are typically disk-dominated, while those with higher densities correspond to bulge-dominated systems \citep{Kauffmann2006}, which is consistent with our sample being mainly blue disks. 
Panel (c) shows that \HI\ detections are concentrated along the star-forming main sequence (SFMS; adopted from \citet{Saintonge2022}), while some massive galaxies with lower star formation rates still have detectable \HI\ gas. 
Finally, panel (d) presents the stellar mass-size relation, where larger galaxies are more likely to have detectable \HI.

The HD2 pilot survey covers a wide range of optical parameter space, for example, including red and low-sSFR galaxies that are typically regarded as quenched systems. In Appendix~\ref{sec:optspec}, we present two examples detected in \HI: a non-confused early-type galaxy (HGC 115430027, $\mathrm{NUV}-r=5.8$, $\log\mathrm{sSFR}/\mathrm{yr}^{-1}=-12.7$) and an interacting galaxy pair with confusion (HGC 115430011, $\mathrm{NUV}-r=4.7$, $\log\mathrm{sSFR}/\mathrm{yr}^{-1} = -13$). Although both sources are characterized by their red colors and very low sSFR, they exhibit differences in morphology and environment.
These examples illustrate the diversity of the \HD2\ sample in the low-sSFR regime.

\section{\HI\ Scaling Relations} \label{sec:scaling}

\subsection{\HI\ Gas Fraction Scaling Relations}

\begin{figure*}[ht!]
\plottwo{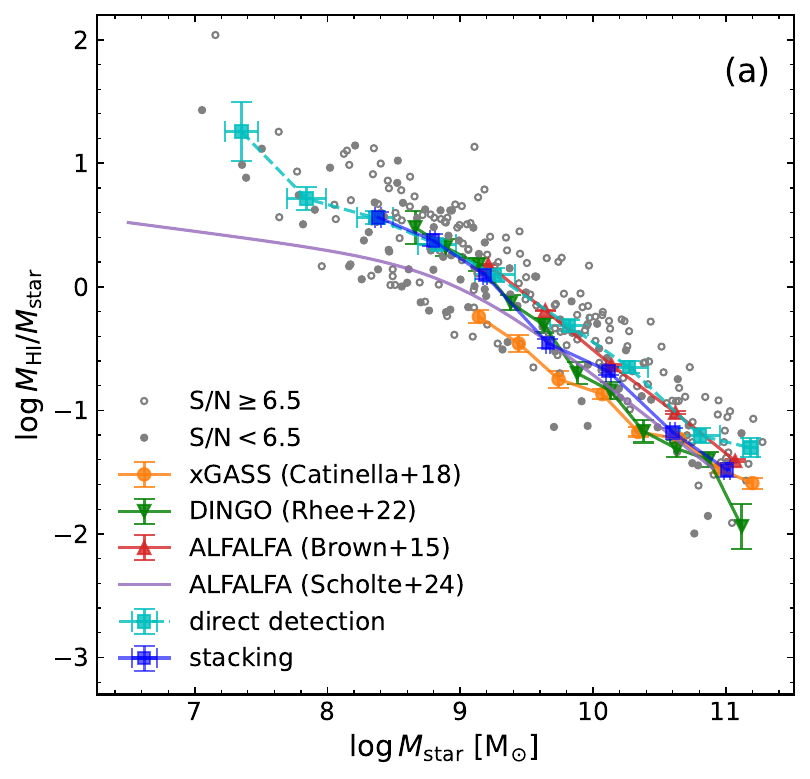}{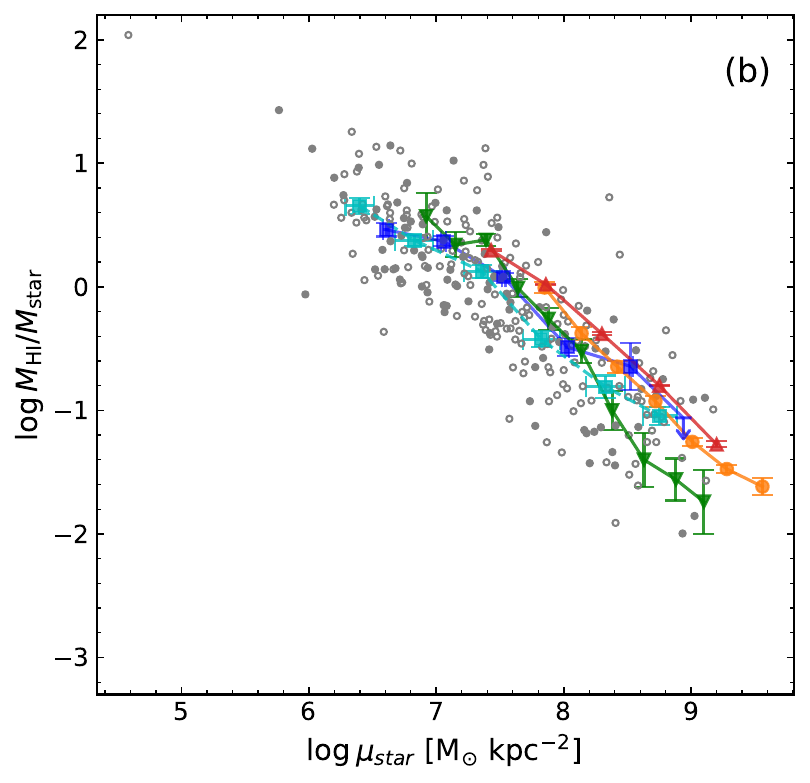}
\plottwo{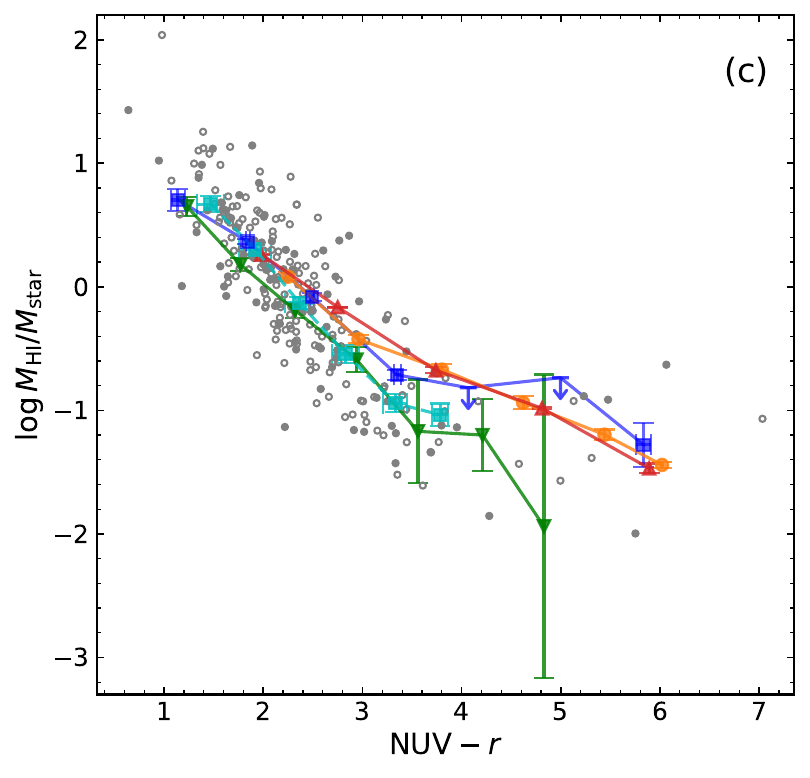}{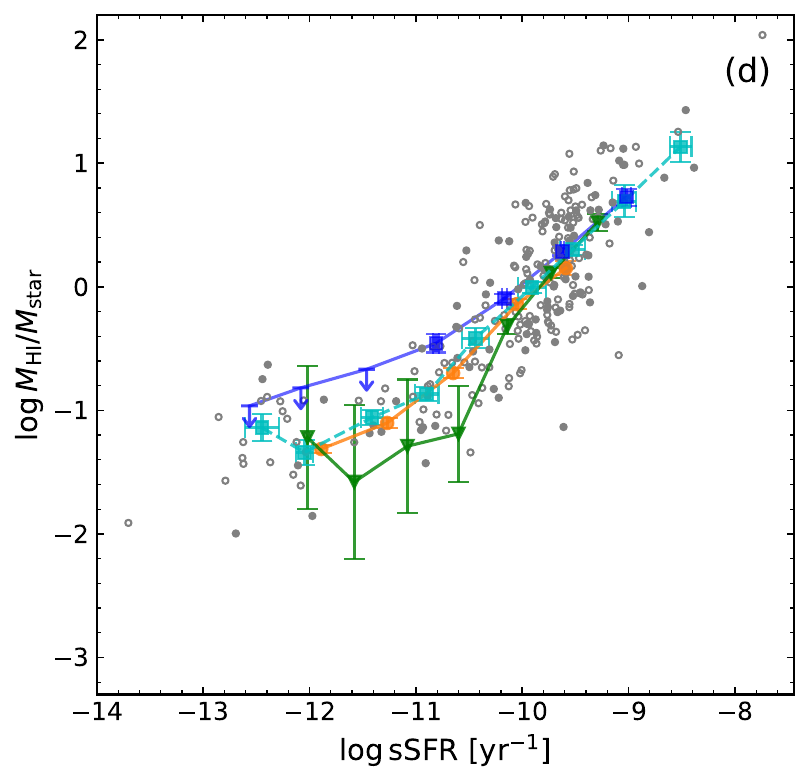}
\caption{The \HI\ gas fraction scaling relations as a function of (a) stellar mass, (b) stellar mass surface density, (c) $\mathrm{NUV}-r$, and (d) specific star formation rate.
Gray dots represent direct detections, while filled and open markers distinguish between high-S/N samples (S/N $\geq$ 6.5) and low-S/N samples (S/N $<$ 6.5).
The lines with square markers are the results of this paper, with direct detection averages shown in cyan dashed lines and stacking results in blue solid lines, with blue downward arrows indicating the upper limits for undetected bins.
We compare them with previous studies, xGASS \citep[orange circles]{Catinella2018} and stacking based on DINGO \citep[green downward triangles]{Rhee2022} and ALFALFA \citep[red upward triangles]{Brown2015}.
In panel (a), the purple solid line shows the mass-complete median stacking results from \citet{Scholte2024}, based on galaxies using ALFALFA and DESI redshifts.
}
\label{fig:frac}
\end{figure*}

To study the cold gas content in relation to galaxy properties, we examine the \HI\ gas fraction scaling relations and compare them with results from other surveys. 
Figure~\ref{fig:frac} shows how the \HI\ gas fraction varies with (a) stellar mass ($M_{\star}$), (b) stellar mass surface density ($\mu_{\star}$), (c) $\mathrm{NUV}-r$, and (d) sSFR.
In each panel of Figure~\ref{fig:frac}, direct detections are shown as gray points, with high-S/N ($\geq 6.5$) in filled and low-S/N ($<$6.5) in open symbols. We select 257 \HI\ detections without confusion (defined in Section~\ref{sec:confuse}) for analysis. Cyan dashed lines with square markers indicate the averages of these detections, while blue lines represent stacking measurements in each bin. 
For direct detections, we report $\langle \log f_{\mathrm{HI}} \rangle$, which is the average of the logarithmic gas fraction since the distribution of \HI\ gas fraction is closer to lognormal than Gaussian \citep{Cortese2011, Brown2015}. 
However, the stacking method only returns another form, $\log \langle f_{\mathrm{HI}} \rangle$, the logarithm of the average. These two statistics are not directly equivalent; the stacking results are typically higher than the direct detection averages at fixed stellar mass (see e.g., Fig.~3 of \citet{Saintonge2022}).
Vertical error bars are estimated using the bootstrap method, while horizontal bars indicate the standard errors of the sample properties. 
The stacking results are also listed in Table~\ref{tab:stack_frac}. 
For comparison, we also include the mean values of the xGASS representative sample \citep{Catinella2018} and two spectral stacking results, from DINGO \citep{Rhee2022} and ALFALFA \citep{Brown2015}.

\begin{deluxetable}{lccccc}
\tabletypesize{\scriptsize} 
\tablewidth{1pt} 
\setlength{\tabcolsep}{10pt}

\tablecaption{The average \HI\ gas fractions for physical properties ($\log M_{\star}$, $\log \mu_{\star}$, $\mathrm{NUV}-r$, $\log \mathrm{sSFR}$) based on the spectral stacking, as shown in Figure~\ref{fig:frac}. $N$ is the number of galaxies stacked in each bin.
The $\mathrm{S/N}_\mathrm{peak}$ is the peak signal-to-noise ratio, defined as the peak flux divided by the noise after stacking.
Gas fraction values with $\mathrm{S/N}_\mathrm{peak} < 3$ are reported as upper limits, with their flux estimated as $3\sigma_\mathrm{rms} \times 200~\mathrm{km~s^{-1}}$. These values are preceded by ``$<$''.
}
\label{tab:stack_frac}

\tablehead{
\colhead{$x$} & \colhead{$\langle x \rangle$} & \colhead{$N$} & \colhead{$\log \langle M_{\mathrm{HI}}/M_{\star} \rangle$} & \colhead{$\mathrm{S/N}_\mathrm{peak}$}
} 
\startdata 
$\log M_{\star}/\mathrm{M}_{\odot}$ & 8.38 & 35 & 0.56 $\pm$ 0.04 & 7.8 \\
 & 8.79 & 55 & 0.38 $\pm$ 0.05 & 9.6 \\
 & 9.19 & 62 & 0.10 $\pm$ 0.03 & 12.1 \\
 & 9.66 & 68 & -0.46 $\pm$ 0.04 & 7.2 \\
 & 10.12 & 59 & -0.68 $\pm$ 0.03 & 13.6 \\
 & 10.60 & 46 & -1.18 $\pm$ 0.03 & 7.3 \\
 & 11.01 & 29 & -1.48 $\pm$ 0.06 & 7.0 \\
\hline
$\log \mu_{\star}/(\mathrm{M}_{\odot}~\mathrm{kpc}^{-2})$ & 6.61 & 38 & 0.46 $\pm$ 0.05 & 11.1 \\
 & 7.06 & 57 & 0.37 $\pm$ 0.04 & 9.4 \\
 & 7.53 & 86 & 0.08 $\pm$ 0.03 & 10.0 \\
 & 8.03 & 72 & -0.49 $\pm$ 0.07 & 5.2 \\
 & 8.53 & 67 & -0.64 $\pm$ 0.19 & 5.9 \\
 & 8.94 & 34 & $<$-1.06 $\pm$ 0.21 & - \\
\hline
$\mathrm{NUV-r}$ & 1.14 & 11 & 0.70 $\pm$ 0.09 & 7.9 \\
 & 1.84 & 97 & 0.36 $\pm$ 0.02 & 12.0 \\
 & 2.49 & 95 & -0.08 $\pm$ 0.03 & 10.3 \\
 & 3.35 & 45 & -0.71 $\pm$ 0.04 & 4.7 \\
 & 4.07 & 18 & $<$-0.82 $\pm$ 0.13 & - \\
 & 5.00 & 12 & $<$-0.74 $\pm$ 0.26 & - \\
 & 5.83 & 14 & -1.28 $\pm$ 0.18 & 3.5 \\
\hline
$\log \mathrm{sSFR}/\mathrm{yr}^{-1}$ & -12.57 & 33 & $<$-0.96 $\pm$ 0.24 & - \\
 & -12.08 & 43 & $<$-0.82 $\pm$ 0.13 & - \\
 & -11.46 & 25 & $<$-0.67 $\pm$ 0.14 & - \\
 & -10.81 & 39 & -0.46 $\pm$ 0.07 & 4.5 \\
 & -10.17 & 72 & -0.09 $\pm$ 0.04 & 10.3 \\
 & -9.62 & 128 & 0.29 $\pm$ 0.03 & 10.5 \\
 & -9.02 & 14 & 0.73 $\pm$ 0.07 & 7.0 \\
\hline
\hline
\enddata

\end{deluxetable}

Figure~\ref{fig:frac}(a) shows a decline in \HI\ fraction with increasing stellar mass. Our direct detections are in good agreement with the ALFALFA stacking trend. As expected, the FAST stacking results are slightly lower than our direct detections for galaxies with $M_{\star} > 10^{9}$ \Msun\, due to the presence of many low S/N and \HI\ non-detected galaxies, which decreases the average fraction.
For the massive end between $10^{9.5}$ and $10^{11}$ \Msun, our stacking results are consistent with those from xGASS and DINGO. 
We also compare with the mass-complete stacking results by \citet{Scholte2024}, who stacked $\sim$70k galaxies using ALFALFA and DESI redshifts. Our mean stacking results are broadly consistent with their median stacking at the massive end, while  \citet{Scholte2024} find a clear flattening of the relation around  $\sim10^{9}$ \Msun  that is not evident in our dataset. 
This difference likely reflects the much larger sample size of \citet{Scholte2024} at low stellar masses, providing robust constraints in the dwarf galaxy regime; our pilot survey does not span sufficient dynamic range to reveal such a feature, but the full \HD2\ survey will increase the number of sources at low stellar masses.

Figure~\ref{fig:frac}(b) shows the \HI\ fraction versus stellar mass surface density. 
All surveys reveal a decreasing trend with $\mu_{\star}$, though we note that the definition of $R_\mathrm{e}$ is not the same among those studies, and it may cause some differences. 
DESI uses a model-based half-light radius combined with the $g, r, z$ bands, whereas xGASS \citep{Catinella2010} and \citet{Brown2015} adopt the $z$-band Petrosian half-light radius, and DINGO derives it from stacked $r+z$ images with axial-ratio corrections \citep{Rhee2022}. But as shown in our previous definition, ours and the other two studies do not consider the ellipticity. 

Figure~\ref{fig:frac}(c) shows a good correlation between $\mathrm{NUV}-r$ and \HI\ fraction.    After excluding galaxies with invalid NUV magnitude, 240 detections remain. On the blue side, our direct detections and stacking results closely align with those from DINGO, while the results from Arecibo \citep{Brown2015, Catinella2018} are systematically higher for redder colors. 
Additionally, the observed flat trend is mainly due to the limitations in detecting the \HI\ fraction and few red galaxies in the low-$z$ sample, consisting of only two upper limits and one detected bin at the red end of our stacking.

Figure~\ref{fig:frac}(d) presents the relation with sSFR. Our direct detections are consistent with the xGASS observations and align with the high-sSFR end of the DINGO survey.
In the range $-13<\log \mathrm{sSFR}/\mathrm{yr^{-1}} <-11$, our direct detections and DINGO also show a flat trend as they approach the detection limit. 
The \HI\ fraction limit in our sample is approximately 1\%, while more massive quiescent galaxies fall below this detection limit \citep{Li2025a}.
And due to the scarcity of low-SFR galaxies, only four detected bins remain in the stacking results.

Overall, these scaling relations support the picture that blue, disk-dominated galaxies contain larger gas reservoirs and higher star formation. With wider survey coverage and larger samples, we will better constrain these trends and trace the process of \HI\ depletion and star formation.

\subsection{\HI\ -- Halo Mass Relation}

\begin{figure}[ht!]
\plotone{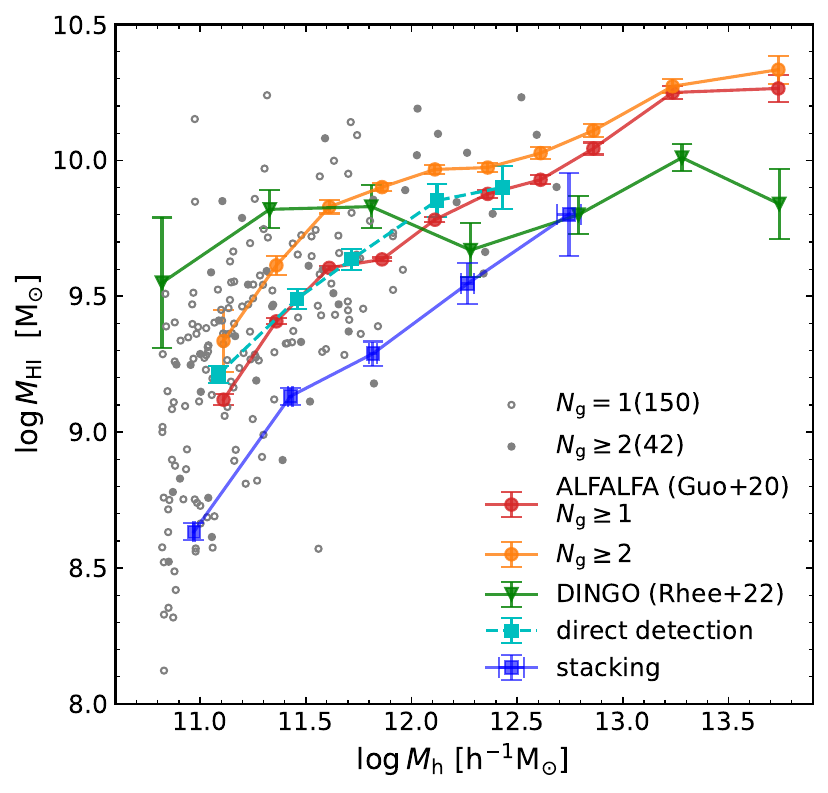}
\caption{The \HI\ -- halo mass relation.
Gray dots represent groups with direct \HI\ detections, with open symbols for $N_g=1$ and filled symbols for $N_g \geq 2$. The cyan dashed line shows the running mean values of direct detections, while blue squares mark the average \HI\ mass from our group stacking. 
We also include ALFALFA stacking from \citet[red circles for $N_g \geq 1$, orange circles for $N_g \geq 2$]{Guo2020} and DINGO group stacking \citep[green downward triangles]{Rhee2022} for comparison.
}
\label{fig:Mhalo-Mhi}
\end{figure}

The \HI\ content is expected to correlate with halo mass, because \HI\ is an effective tracer of dark matter. 
In the \HI\ -- halo mass relation, we do not separate centrals and satellites like \citet{Yan2025} owing to confusion from FAST's large beam and limited volume. Instead, we co-add all flux within the halo radius or beam size. 
Figure~\ref{fig:Mhalo-Mhi} shows the \HI\ -- halo mass relation, with stacking results listed in Table~\ref{tab:stack_Mhalo}.
Groups with detectable \HI\ emission are shown as gray dots, with open and filled symbols distinguishing isolated systems ($N_g = 1$) from groups ($N_g \geq 2$). Most of our \HI-detected halos have $N_g = 1$, while only a few with $N_g \geq 2$ appear in the $10^{12} - 10^{13}~\mathrm{h^{-1}}$\Msun\ range. 
The cyan dashed lines indicate the mean values of these groups. Our direct detections are dominated by isolated galaxies with halo masses below $10^{12}~\mathrm{h^{-1}}$~\Msun, which is natural for \HI-detected populations. 
For comparison, the red and orange lines from \citet{Guo2020}, based on the stacking of ALFALFA groups, illustrate that halos with a higher number of members contain more \HI. 
Our direct detections show good agreement with the group stacking results from ALFALFA when selecting $N_g \geq 1$ \citep{Guo2020}.
The green line with downward triangles represents the results from the early science phase of DINGO \citep{Rhee2022}, which involved stacking both group central and satellite galaxies, revealing a nearly flat curve across all mass ranges.
The blue lines with square markers illustrate the stacking results of the \HD2\ pilot survey, displaying a lower \HI\ mass compared to other works. The highest mass bins are associated with large uncertainties due to the limited number of massive halos observed in 10 \deg2.

Previous studies find that the \HI\ -- halo mass relation rises with halo mass at the low-mass end and then flattens into a plateau \citep[e.g.][]{Guo2021, Saraf2024}. 
Some semi-analytical models (e.g., \texttt{GALFORM}, \citealt{Baugh2019}; \texttt{SHARK}, \citealt{Chauhan2020}) predict a dip at $10^{11.8}-10^{13}$ \Msun\ due to virial shock heating and AGN feedback \citep{Guo2020}. 
However, this feature is not evident in current observations or in other models such as TNG100 \citep{Villaescusa-Navarro2018}, \texttt{GAEA} \citep{Spinelli2020}, the \texttt{ELUCID} SAM catalog \citep{Zhang2022b}, and an \HI\ mass estimator calibrated on xGASS \citep{Li2022}.
As our pilot survey includes only a limited number of massive halos, we did not present model predictions in Figure~\ref{fig:Mhalo-Mhi}. But a larger \HD2\ dataset could offer tighter constraints on these models and establish a reliable scaling relation in the future.

\begin{deluxetable}{lcccc}
\tabletypesize{\scriptsize}
\tablewidth{1pt} 
\setlength{\tabcolsep}{13pt}

\tablecaption{The \HI\ -- halo mass relation based on the group stacking, as shown in Figure~\ref{fig:Mhalo-Mhi}. $N$ is the number of halos stacking in each bin.
The $\mathrm{S/N}_\mathrm{peak}$ is the peak S/N, which has the same meaning as Table~\ref{tab:stack_frac}.
}
\label{tab:stack_Mhalo}

\tablehead{
\colhead{$\log M_\mathrm{h}/(\mathrm{h^{-1}M_{\odot}})$} & \colhead{$N$} & \colhead{$\log \langle M_{\mathrm{HI}} \rangle$} & \colhead{$\mathrm{S/N}_\mathrm{peak}$}
} 
\startdata 
10.97 & 474 &  8.63 $\pm$ 0.03 &  8.9 \\
11.43 & 136 &  9.13 $\pm$ 0.03 &  10.4 \\
11.82 & 44 &  9.29 $\pm$ 0.04 &  6.7 \\
12.27 & 19 &  9.55 $\pm$ 0.08 &  6.0 \\
12.75 & 5 &  9.80 $\pm$ 0.15 &  5.0 \\
\enddata
\end{deluxetable}

\section{Summary}\label{sec:summary}

We conduct a 10 deg$^{2}$ pilot survey in a region overlapping with HectoMAP, HSC-SSP, and DESI EDR, serving as a precursor to the ongoing \HD2\ project. The data are processed using \HIFAST\ \citep{Jing2024}, a calibration and imaging pipeline for FAST \HI\ observations.
We detect 339 \HI\ sources at $z < 0.09$, measure their physical parameters, and perform \HI\ stacking experiments based on DESI EDR spectroscopic redshifts.
Our main results are summarized as follows:

\begin{enumerate}
    \item At $z<0.09$, we detect 339 \HI\ sources: 259 with \SOFIA\ and the rest confirmed with DESI positions and redshifts. With an integration time of 7.3 minutes per beam, the pilot survey achieves a detection sensitivity of 0.45~m\Jybeam\ at 4.8~\kms\ resolution and a source number density of 33.9 deg$^{-2}$. 
    
    \item DESI spectroscopic redshifts greatly improve optical counterpart matching when compared with photo-$z$ and only photometry surveys. For galaxies with apparent magnitude $r < 19.5$, both the matching and correct rates exceed 90\% using DESI EDR SV3 spectra, successfully improved compared to SDSS.
    We also quantify the significant confusion fraction as 20\% in the pilot survey using DESI redshifts. This higher rate arises because DESI can probe fainter and smaller galaxies, while the confusion rate based on SDSS redshifts would be underestimated.

    \item In the DESI EDR BGS bright samples, approximately 50\% of galaxies exhibit \HI\ detections when constraining the magnitude to $r < 17.8$ mag within the redshift range $0.01 < z < 0.05$, utilizing an on-source integration time comparable to that of xGASS. 
    We also investigate the distribution of \HI\ detections and non-detections within the optical view. The \HI\ detections encompass nearly the entire range of the DESI BGS sample.

    \item With the high completeness of DESI redshifts, we derive \HI\ fraction scaling relations with stellar mass, stellar mass surface density, $\mathrm{NUV}-r$, and sSFR.
    The pilot survey results are consistent with previous detections \citep{Catinella2018} and stacking studies \citep{Brown2015, Rhee2022}. 

    \item We also attempt to revisit the \HI\ -- halo mass relation and find our direct detections agree well with the group stacking of ALFALFA \citep{Guo2020}. Due to the small survey area, few massive halos are included, preventing us from distinguishing between the flat trend and the shallow dip predicted in some simulations within $10^{11.8}-10^{13}$ \Msun.
    
\end{enumerate}

As the pilot survey of the ongoing \HD2\ program,  this 10 \deg2 field primarily provides valuable experience for the full survey.
We first demonstrate the improved correctness of optical counterpart matching when using DESI EDR compared to SDSS, highlighting the importance of DESI spectra in guiding \HI\ observations.
Although the pilot survey is shallower than the final \HD2\ sensitivity and the scaling relations obtained are easily influenced by cosmic variance and selection effects, it confirms that most technical challenges have been addressed and demonstrates the scientific potential of the full survey.

The full \HD2\ survey will cover a contiguous field centered on the Boötes void of a few hundred square degrees, and will achieve a detection sensitivity of 0.28~m\Jybeam\ at a resolution of 4.8~\kms.
We expect a source detection rate of more than 30 sources per deg$^2$ and an \HI\ detection rate exceeding 60\% for DESI BGS selected galaxies at $z < 0.09$.
At present, Year 1 and Year 2 observations have already been completed, covering the 100 \deg2 at a depth equivalent to the pilot survey and the first data release of the \HD2\ survey will be based on these observations.

Building on these observational capabilities, the \HD2\ survey is mainly positioned to deliver two major scientific advantages: a high-completeness \HI\ census of the local Universe enabled by DESI's spectroscopic depth, and a large-scale mapping of the \HI\ gaseous environment enabled by its contiguous sky coverage. 
For local \HI\ science, the combination of FAST sensitivity and DESI spectroscopy will enable studies of scaling relations along the atomic gas sequence, the \HI\ mass and velocity width functions, the Tully--Fisher relation, and the cosmic \HI\ mass density, while the deep HSC-SSP imaging will allow identification of rare populations such as RELHIC candidates \citep{Benitez-Llambay2023} and low surface brightness galaxies. 
For large-scale structure, the contiguous footprint makes the \HD2\ survey suited to probing the dependence of \HI\ content on large-scale environment and cosmic web structure, complementing ongoing medium-area interferometric surveys such as DINGO and MIGHTEE-HI with significantly higher single-dish sensitivity. 
We anticipate that these deep \HI\ data will provide insights into the connection between atomic gas reservoirs and galaxy properties in the local Universe.

\begin{acknowledgments}
We would like to thank the referee and editors for their constructive suggestions and comments. 
We thank Martha P. Haynes for her insightful discussions on source flux calculations and parameter measurements. 
We also extend our gratitude to Simon D. M. White for his thoughtful suggestions on various aspects of our study. 
Additionally, we appreciate the valuable discussions with Yinghui Zheng, Chuan-Peng Zhang, Hengxing Pan, Ruilei Zhou, Fujia Li, and Lin Du regarding data reduction and \HI\ sciences. 

This work was supported by the China National Key Program for Science and Technology Research and Development of China (No. 2022YFA1602901), the National SKA Program of China (No. 2022SKA0110201), the National Natural Science Foundation of China (NSFC; Nos. 12125302, 12588202), the CAS Project for Young Scientists in Basic Research grant (No. YSBR-062), and the science research grants from the China Manned Space Project. 
Y.J. acknowledges support from the Cultivation Project for FAST Scientific Payoff and Research Achievement of CAMS-CAS.
H.Z. acknowledges the supports from National Key R\&D Program of China (Nos. 2022YFA1602902 and 2025YFE0202300), and NSFC (Nos. 12120101003 and 12373010).
T.L. is supported  by the  Fundamental Research  Funds  for  the  Central  Universities (2026MS217).
N.Y. is supported by the projects funded by China Postdoctoral Science Foundation No. 2022M723175 and GZB20230766.
H.G. is supported by NSFC No. 12595313 and CAS Project for Young Scientists in Basic Research (YSBR-092).
H.H. is supported by the NSFC No. 12503012.
T.F. is supported by the National SKA Program of China (No. 2025SKA0150103), NSFC (Nos. 12550002, 12133008, 12221003, 11890692), and acknowledges the science research grants from the China Manned Space Project (Nos. CMS-CSST-2021-A04 and CMS-CSST-2025-A10).
J.P. acknowledges support from the National Key R\&D Program of China (No. 2018YFE0202900).

This work made use of the data from FAST (Five-hundred-meter Aperture Spherical radio Telescope)(\url{https://cstr.cn/31116.02.FAST}). FAST is a Chinese national mega-science facility, operated by National Astronomical Observatories, Chinese Academy of Sciences.

The DESI Legacy Imaging Surveys consist of three individual and complementary projects: the Dark Energy Camera Legacy Survey (DECaLS), the Beijing-Arizona Sky Survey (BASS), and the Mayall z-band Legacy Survey (MzLS). 
DECaLS, BASS and MzLS together include data obtained, respectively, at the Blanco telescope, Cerro Tololo Inter-American Observatory, NSF's NOIRLab; the Bok telescope, Steward Observatory, University of Arizona; and the Mayall telescope, Kitt Peak National Observatory, NOIRLab. NOIRLab is operated by the Association of Universities for Research in Astronomy (AURA) under a cooperative agreement with the National Science Foundation. 
Pipeline processing and analyses of the data were supported by NOIRLab and the Lawrence Berkeley National Laboratory (LBNL). 
Legacy Surveys also uses data products from the Near-Earth Object Wide-field Infrared Survey Explorer (NEOWISE), a project of the Jet Propulsion Laboratory/California Institute of Technology, funded by the National Aeronautics and Space Administration. 
Legacy Surveys was supported by: the Director, Office of Science, Office of High Energy Physics of the U.S. Department of Energy; the National Energy Research Scientific Computing Center, a DOE Office of Science User Facility; the U.S. National Science Foundation, Division of Astronomical Sciences; the National Astronomical Observatories of China, the Chinese Academy of Sciences and the Chinese National Natural Science Foundation. LBNL is managed by the Regents of the University of California under contract to the U.S. Department of Energy.
The complete acknowledgments can be found at \url{https://www.legacysurvey.org/acknowledgment/}.
The Photometric Redshifts for the Legacy Surveys (PRLS) catalog used in this paper was produced thanks to funding from the U.S. Department of Energy Office of Science, Office of High Energy Physics via grant DE-SC0007914.

Funding for the Sloan Digital Sky Survey IV has been provided by the Alfred P. Sloan Foundation, the U.S. Department of Energy Office of Science, and the Participating Institutions.
SDSS-IV acknowledges support and resources from the Center for High Performance Computing at the University of Utah. The SDSS website is \url{https://www.sdss4.org}.
SDSS-IV is managed by the Astrophysical Research Consortium for the Participating Institutions of the SDSS Collaboration including the Brazilian Participation Group, the Carnegie Institution for Science, Carnegie Mellon University, Center for Astrophysics | Harvard \& Smithsonian, the Chilean Participation Group, the French Participation Group, Instituto de Astrof\'isica de Canarias, The Johns Hopkins University, Kavli Institute for the Physics and Mathematics of the Universe (IPMU)/University of Tokyo, the Korean Participation Group, Lawrence Berkeley National Laboratory, Leibniz Institut f\"ur Astrophysik Potsdam (AIP), Max-Planck-Institut f\"ur Astronomie (MPIA Heidelberg), Max-Planck-Institut f\"ur Astrophysik (MPA Garching), Max-Planck-Institut f\"ur Extraterrestrische Physik (MPE), National Astronomical Observatories of China, New Mexico State University, New York University, University of Notre Dame, Observat\'orio Nacional/MCTI, The Ohio State University, Pennsylvania State University, Shanghai Astronomical Observatory, United Kingdom Participation Group, Universidad Nacional Aut\'onoma de M\'exico, University of Arizona, University of Colorado Boulder, University of Oxford, University of Portsmouth, University of Utah, University of Virginia, University of Washington, University of Wisconsin, Vanderbilt University, and Yale University.

The Hyper Suprime-Cam (HSC) collaboration, comprising the astronomical communities of Japan and Taiwan as well as Princeton University, has seen the development of its instrumentation and software by the National Astronomical Observatory of Japan (NAOJ), the Kavli Institute for the Physics and Mathematics of the Universe (Kavli IPMU), the University of Tokyo, the High Energy Accelerator Research Organization (KEK), the Academia Sinica Institute for Astronomy and Astrophysics in Taiwan (ASIAA), and Princeton University. Funding for this endeavor has been provided by a variety of sources, including the FIRST program from the Japanese Cabinet Office, the Ministry of Education, Culture, Sports, Science and Technology (MEXT), the Japan Society for the Promotion of Science (JSPS), the Japan Science and Technology Agency (JST), the Toray Science Foundation, NAOJ, Kavli IPMU, KEK, ASIAA, and Princeton University.

\textit{GALEX} is a NASA Small Explorer, launched in April 2003. It acknowledges NASA's support for construction, operation, and science analysis for the \textit{GALEX} mission, developed in cooperation with the Centre National d'Etudes Spatiales (CNES) of France and the Korean Ministry of Science and Technology.

We also made use of the NASA/IPAC Extragalactic Database (NED), which is funded by the National Aeronautics and Space Administration and operated by the California Institute of Technology; ``Aladin sky atlas" and \texttt{ipyaladin} developed at CDS, Strasbourg Observatory, France.

\end{acknowledgments}

\vspace{5mm}
\facilities{DESI, FAST, \textit{GALEX}, SDSS, Subaru }

\software{Aladin Lite v3 \citep{Baumann2022}, Astropy \citep{TheAstropyCollaboration2022}, \HIFAST\ \citep{Jing2024}, HiPS\footnote{Hierarchical Progressive Surveys: \url{https://aladin.cds.unistra.fr/hips/}} \citep{Fernique2015}, 
ipyaladin \citep{Boch2020}, kcorrect \citep{Blanton2007}, 
NumPy \citep{Harris2020}, petrofit \citep{Geda2022}, photutils \citep{Bradley2023}, SciPy\citep{Virtanen2020},  \SOFIA\ \citep{Serra2015, Westmeier2021} 
}

\appendix

\section{The \HI\ Extragalactic Source Catalog} \label{sec:catalog}

Here we present the first 5 rows of the \HI\ extragalactic source catalog of the pilot \HD2\ survey, followed by the description of each column.

\begin{splitdeluxetable*}{lcccccccBccccccccccc}
\tabletypesize{\scriptsize}
\tablewidth{0pt} 
\tablecaption{The \HD2\ Pilot Survey Extragalactic \HI\ Source Catalog}
\label{tab:catalog}

\tablehead{
\colhead{HGC ID} & \colhead{HI Position} & \colhead{OC Position} & \colhead{$V_{\mathrm{HI}}$} & \colhead{$V_{\mathrm{OC}}$} & \colhead{$z_\mathrm{CMB}$} & \colhead{$W_{\mathrm{50}}$} & \colhead{$W_{\mathrm{20}}$} & \colhead{$S$} & \colhead{$\sigma_\mathrm{rms}$} & \colhead{S/N} & \colhead{$V_{\mathrm{85}}^{\mathrm{cog}}$} & \colhead{$S^{\mathrm{cog}}$} & \colhead{S/N$^{\mathrm{cog}}$} & \colhead{$D$} & \colhead{$\log M_{\mathrm{HI}}$} & \colhead{$P_{\mathrm{bestOC}}$} & \colhead{HI code} & \colhead{OC code} 
\\
\colhead{} & \colhead{(J2000)} & \colhead{(J2000)} & \colhead{($\mathrm{km~s^{-1}}$)} & \colhead{($\mathrm{km~s^{-1}}$)} & \colhead{} & \colhead{($\mathrm{km~s^{-1}}$)} & \colhead{($\mathrm{km~s^{-1}}$)} & \colhead{($\mathrm{Jy~km~s^{-1}}$)} & \colhead{($\mathrm{mJy}$)} & \colhead{} & \colhead{($\mathrm{km~s^{-1}}$)} & \colhead{($\mathrm{Jy~km~s^{-1}}$)} & \colhead{} & \colhead{(Mpc)} & \colhead{($\log~M_{\odot}$)} & \colhead{} & \colhead{} & \colhead{}
} 
\colnumbers
\startdata 
115430000 & 153621.2 +434048 & 153621.40 +433845.3 & 14990 & 15035 & 0.0503 & 173(45) & 228 & 0.299(0.031) & 0.65 & 7.8 & 208(4) & 0.286(0.007) & 6.8 & 199.4 & 9.4(0.12) & 0.12 & G & C \\
9930 & 153626.4 +433122 & 153627.02 +433107.5 & 6096 & 5981 & 0.0206 & 293(10) & 417 & 6.456(0.039) & 0.74 & 114.8 & 403(2) & 6.864(0.011) & 103.7 & 85.2 & 10.03(0.05) & 0.43 & G & G \\
9931 & 153636.0 +432503 & 153635.15 +432528.7 & 5809 & 5798 & 0.0196 & 280(10) & 356 & 3.852(0.038) & 0.75 & 68.2 & 241(2) & 3.871(0.008) & 73.8 & 80.8 & 9.76(0.05) & 0.92 & G & G \\
115430001 & 153639.5 +432830 & 153641.04 +432817.5 & 23016 & 23010 & 0.077 & 134(15) & 154 & 0.233(0.02) & 0.64 & 7.0 & 139(3) & 0.25(0.004) & 7.4 & 334.9 & 9.72(0.1) & 0.29 & G & C \\
9933 & 153642.2 +433203 & 153642.16 +433221.6 & 5584 & 5577 & 0.0189 & 429(37) & 510 & 1.036(0.024) & 0.45 & 23.9 & 361(3) & 0.979(0.005) & 25.5 & 76.2 & 9.14(0.06) & 0.96 & G & G \\
\enddata
\tablecomments{This table is published in its entirety in the machine-readable format.
The first 5 rows are shown here for guidance regarding their form and content. 
More details are also available online at \url{https://ccg-fast.github.io/hd2/data}.}
\end{splitdeluxetable*}

\begin{itemize}

\item \textbf{Column 1.} Unique index number for the Hundred-\deg2 Galaxy Catalog (HGC).
Galaxies numbered 1 -- 12943 correspond to UGC identifiers \citep{Nilson1973}.
For galaxies with a 9-digit identifier, the first digit indicates the Northern hemisphere, the next four digits encode hours of R.A. and degrees of Decl., and the final four digits specify the source within that range.

\item \textbf{Column 2.} Flux-weighted centroid of the \HI\ source in \texttt{hhmmss.s +ddmmss}. Systematic telescope pointing errors (typically $\sim10''$, varying with R.A. and Decl. in OTF mode) have been corrected.

\item \textbf{Column 3.} Centroid of the most probable optical counterpart following automated and manual examinations, in \texttt{hhmmss.ss +ddmmss.s}. See Section~\ref{sec:optcp} for details.

\item \textbf{Column 4.} Heliocentric velocity of the \HI\ source in \kms, following the optical velocity convention:
$V_{\mathrm{HI}} = cz_{\odot} = c(\delta \lambda / \lambda_0)$.
Here $V_{\mathrm{HI}}$ is measured with the method of \citetalias{Springob2005}, defined as the midpoint between the two velocities at which the flux density drops to 50\%. Its uncertainty is taken as half the error in $W_{50}$ (Column 7) \citep{Giovanelli2007}.

\item \textbf{Column 5.} Barycentric velocity of the most probable OC, $V_{\mathrm{OC}}$ in~\kms\ , provided by spectroscopic surveys if the optical redshift is available. 

\item \textbf{Column 6.} \HI\  redshift in the cosmic microwave background (CMB) reference frame, $z_\mathrm{CMB}$, converted from heliocentric redshift $z_{\odot}$ in Column 4 \citep{Fixsen1996, Courteau1999}. 

\item \textbf{Columns 7 and 8.} Velocity width of the \HI\ profile, $W_{50}$ and $W_{20}$ in \kms, measured at the 50\% or 20\% level of the two
peaks using \citetalias{Springob2005}'s method, as described in Section~\ref{sec:width}.
Only instrumental broadening has been corrected.
The uncertainties are dominated by polynomial fitting errors, mostly contributed by their slopes and the spectrum's rms or S/N.

\item \textbf{Column 9.} Integrated \HI\ flux of the source, $S$, in Jy~\kms.
The flux error has been simply estimated by $\sigma_S = \sqrt{N} 
 \sigma_\mathrm{rms} \delta v$, where $N$ is the integrated channel numbers and $\delta v$ is the mean velocity resolution at this redshift.

\item \textbf{Column 10.} Noise level, $\sigma_\mathrm{rms}$ of the spatially integrated line profile in the unit of mJy. It is measured on the RFI-free and signal-free part of the spectrum.
The spectral resolution is about 10~\kms\ after a three-channel down-sample operation and a three-point [0.25, 0.50, 0.25] Hanning smoothing \citep{Springob2005}.

\item \textbf{Column 11.} The S/N has been estimated using the same method as \citet{Haynes2018}, 
\begin{equation}
    \mathrm{S} / \mathrm{N}=\left(\frac{S}{W_{50}}\right) \frac{w_{\mathrm{smo}}^{1 / 2}}{\sigma_{\mathrm{rms}}},
    \label{equ:S/N}
\end{equation}
where $S$ is the integrated flux in Jy~\kms\ listed in Column 9; $S / W_{50}$ is the mean flux density in Jy; $\sigma_{\mathrm{rms}}$ is the rms in Column 10 with the unit of Jy.
As the spectra for parameter measurement have the same spectral resolution as ALFALFA, we adopt the smooth factor $w_{\mathrm{smo}}$ like \citet{Giovanelli2007}, $W_{50}/(2\times10)$ for $W_{50} < 400$~\kms\ or $400/(2\times10)=20$ for $W_{50} \geq 400$ \kms.
Candidates with S/N $<$ 3 are discarded in the final catalog.

\end{itemize}

Columns 12 to 14 are based on \citetalias{Yu2020}'s CoG method:

\begin{itemize}

\item \textbf{Column 12.} Line width $V_{85}^{\mathrm{cog}}$ (\kms), enclosing 85\% of the total flux (see Section~\ref{sec:width}). Uncertainty is estimated by adding noise to the spectrum and taking the standard error of mock realizations.

\item \textbf{Column 13.} Total flux $S^{\mathrm{cog}}$ in Jy~\kms, defined by the flat part of the CoG. Errors are derived with the same procedure as in Column 12 and are generally smaller than those from \citetalias{Springob2005}.

\item \textbf{Column 14} S/N from the CoG method, using $S^{\mathrm{cog}}$ (Col. 13) and $V_{85}^{\mathrm{cog}}$ (Col. 12). Since $V_{85}^{\mathrm{cog}}$ is usually smaller than $W_{50}$, this S/N is slightly higher than that in Column 11.

\end{itemize}

Columns 15 to 19 list the distance, \HI\ mass, probability of the best OC, and category codes:

\begin{itemize}

\item \textbf{Column 15.} Distance in Mpc, derived from the Cosmicflows-4 calculator\footnote{\url{http://edd.ifa.hawaii.edu/CF4calculator/}} \citep{Kourkchi2020, Valade2024}; only three galaxies have Tully-Fisher distances, which we do not adopt.

\item \textbf{Column 16.} Logarithm of the \HI\ mass, $M_{\mathrm{HI}}$ in the solar mass unit, computed as\footnote{All the parameters mentioned in this catalog are consistent with the observed frame, see \citet{Meyer2017} for more details.},
\begin{equation}
    M_{\mathrm{HI}} = \frac{2.35\times 10^5}{(1+z)^2} D^2 S,
\label{equ:himass}
\end{equation}
where $D$ is the distance defined in Column 15 and $S$ is the \HI\ flux provided in Column 9. 
The error of \HI\ mass is estimated following Equation~5 in \citet{Haynes2018}, with a distance uncertainty of 5\% \citep{Zhang2024}. 

\item \textbf{Column 17.} The probability of the best OC, $P_{\mathrm{bestOC}}$, is determined using the method described in Section~\ref{sec:match}.

\item \textbf{Column 18.} The \HI\ source category code: 

\begin{itemize}
    \item \texttt{G} represents `good' S/N $\geq 6.5$ (230 sources); 
    \item \texttt{L} means `low' S/N, $3 \leq $ S/N $< 6.5$ (109 sources).
\end{itemize}

\item \textbf{Column 19.} The optical counterpart category code, see the examples in Figure~\ref{fig:app_optspec3}:

\begin{itemize}
    \item \texttt{G} (good) means that the source has an evident OC with good quality (267 sources);
    \item \texttt{B} (bad) corresponds to   the OC contaminated by a nearby bright foreground star whose diffraction spikes prevent reliable spectroscopic or photometric identification (3 sources);
    \item \texttt{C} (confusion) refers to the case that the \HI\ source has more than one possible OC, as defined in Section~\ref{sec:confuse} (59 sources);
    \item \texttt{U} (uncertain) represents that the spectroscopic redshift is unavailable or DESI provides redshifts with a warning bitmask. So we still flag it with `uncertain' (10 sources).

\end{itemize}

\end{itemize}

\section{Optical Images and \HI\ Spectra of Some Sources }\label{sec:optspec}

\begin{figure*}[ht!]
\centering
\includegraphics[width=1.5\columnwidth]{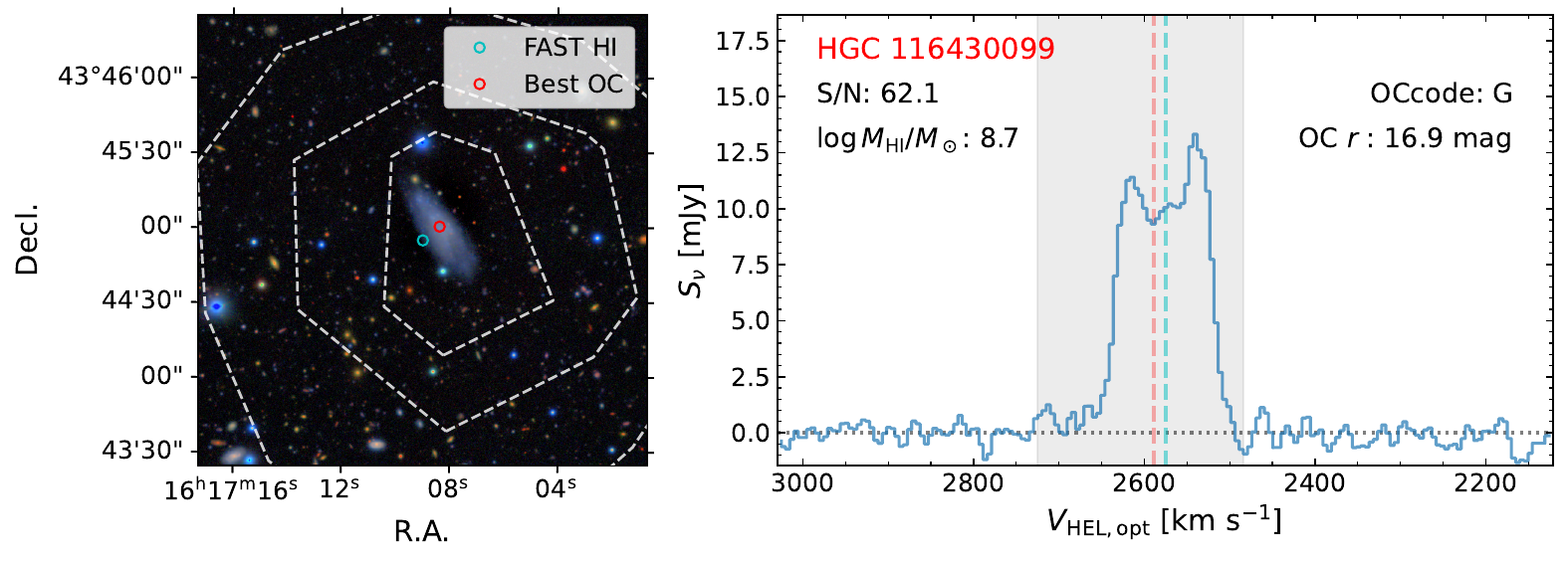}

\includegraphics[width=1.5\columnwidth]{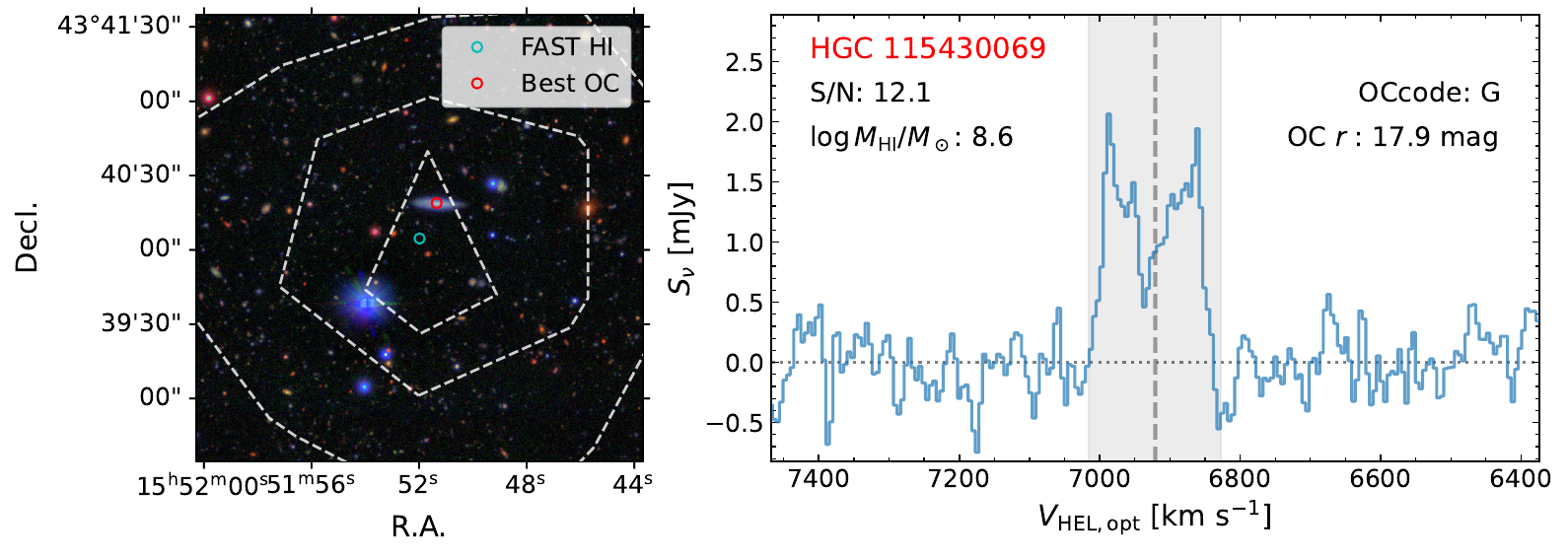}

\includegraphics[width=1.5\columnwidth]{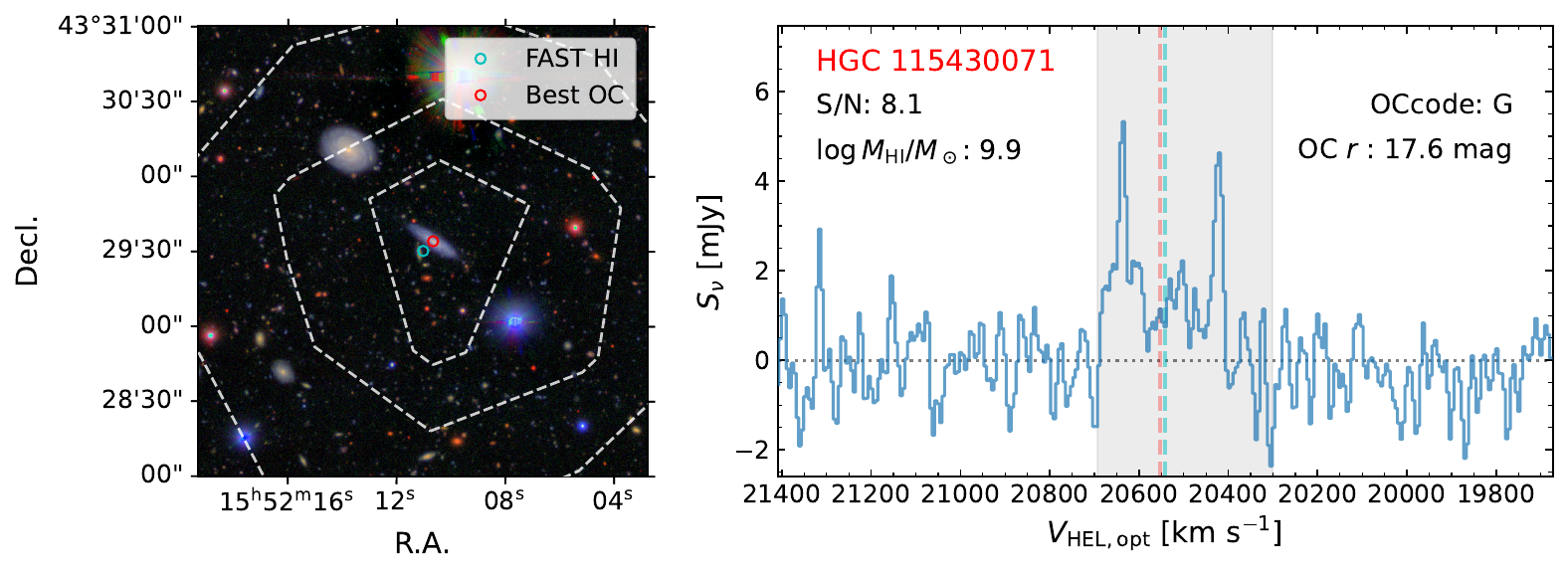}
\caption{Representative \HI\ detections with high S/N. For each source, the left panel shows the HSC-SSP optical image overlaid with \HI\ flux contours (white dashed lines). The cyan circle marks the \HI\ centroid and the red circle marks the best optical counterpart. The right panel shows the \HI\ spectrum, with the red dashed line indicating the optical redshift of the best OC and the cyan dashed line indicating the \HI\ heliocentric velocity. 
The gray shaded region indicates the velocity range integrated to measure the \HI\ flux.
The S/N, $\log M_\mathrm{HI}/M_\odot$, OC code, and $r$-band apparent magnitude of the best OC are labeled in each panel.
We note that the spiral galaxy visible in the upper-left corner of the bottom-row panel (HGC~115430071) is a foreground galaxy at 17940~\kms, offset by $\sim$2600~\kms\ from the \HI\ central velocity of this source.}
\label{fig:app_optspec}
\end{figure*}

\begin{figure*}[ht!]
\centering
\includegraphics[width=1.5\columnwidth]{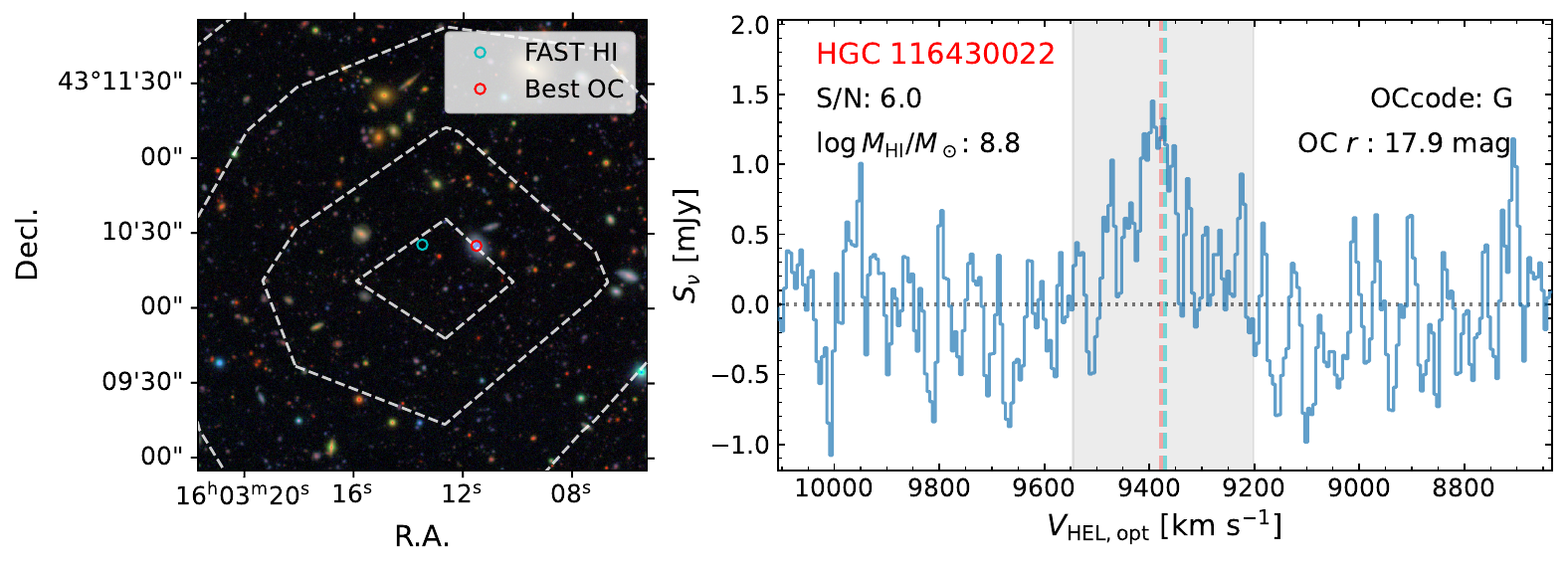}

\includegraphics[width=1.5\columnwidth]{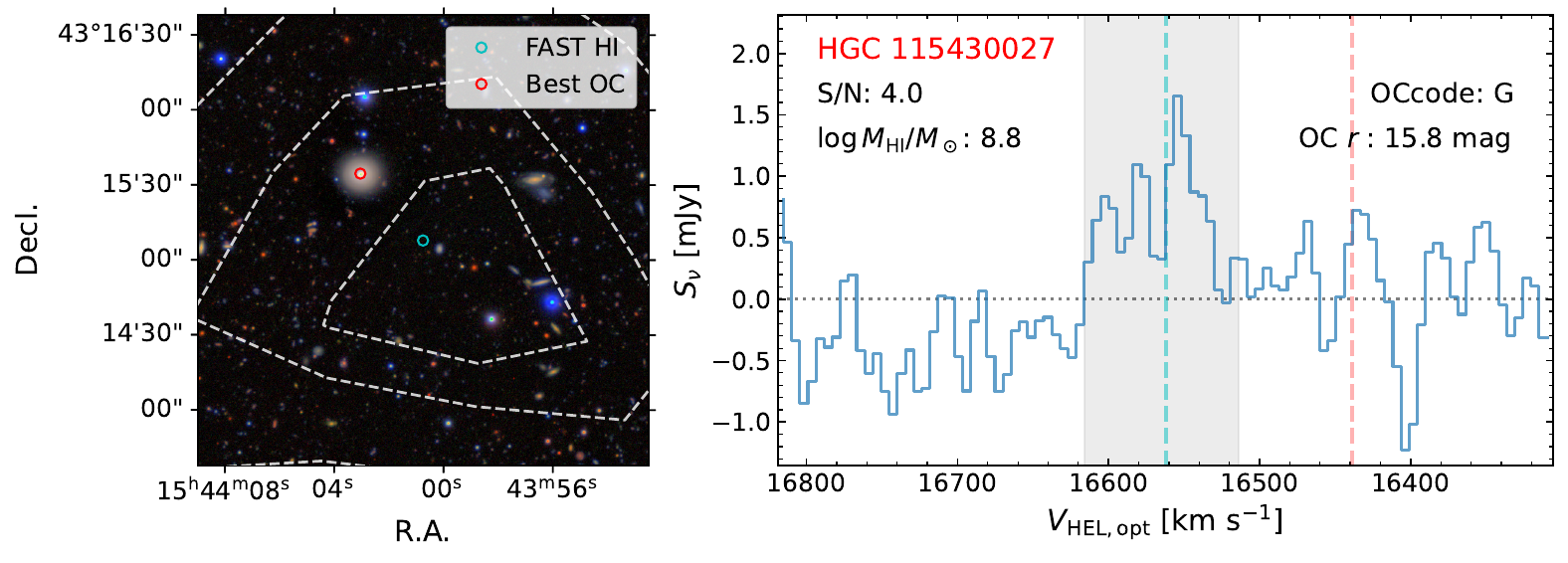}

\includegraphics[width=1.5\columnwidth]{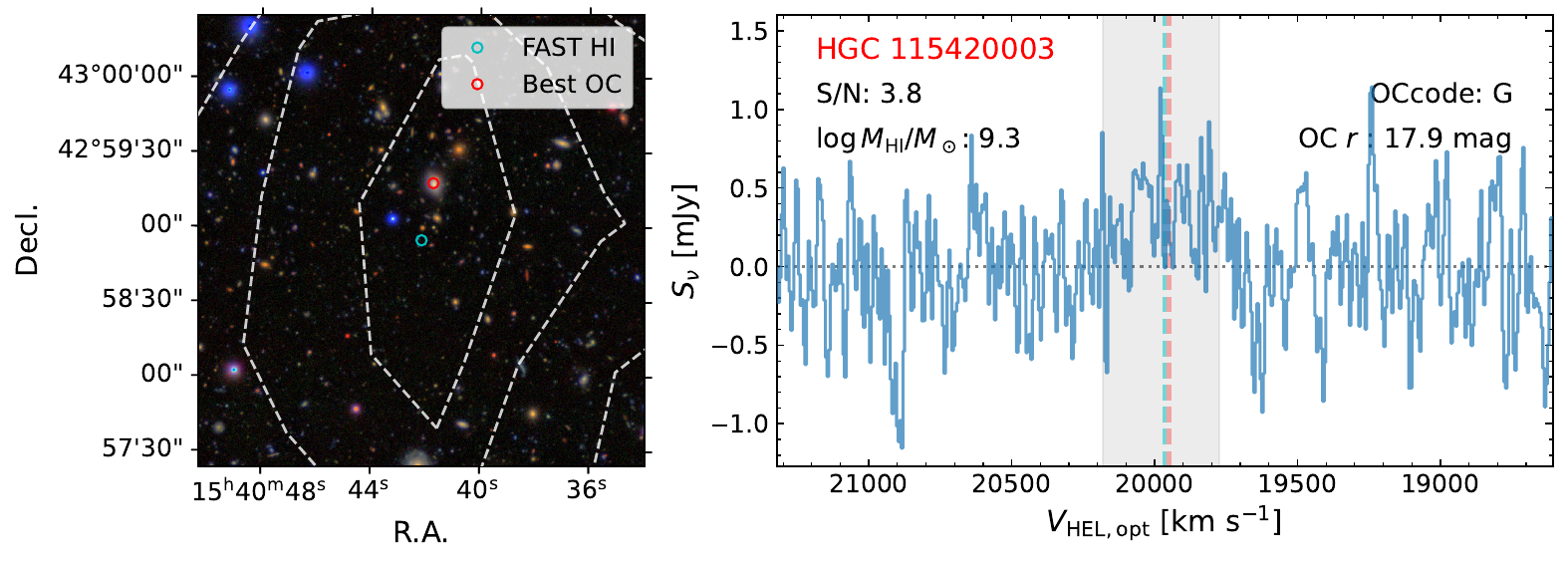}

\caption{Same as Figure~\ref{fig:app_optspec}, but for sources with lower S/N. 
The middle row presents an early-type galaxy characterized by a low sSFR and a red NUV$-r$ color, serving as an example of galaxies that fall below SFMS.
The bottom row is a source recovered through targeted search at a known DESI redshift position, having been initially missed by \SOFIA\ due to its low S/N = 3.8. 
These examples illustrate the data quality of the low S/N samples.}
\label{fig:app_optspec2}
\end{figure*}

\begin{figure*}[ht!]
\centering
\includegraphics[width=1.5\columnwidth]{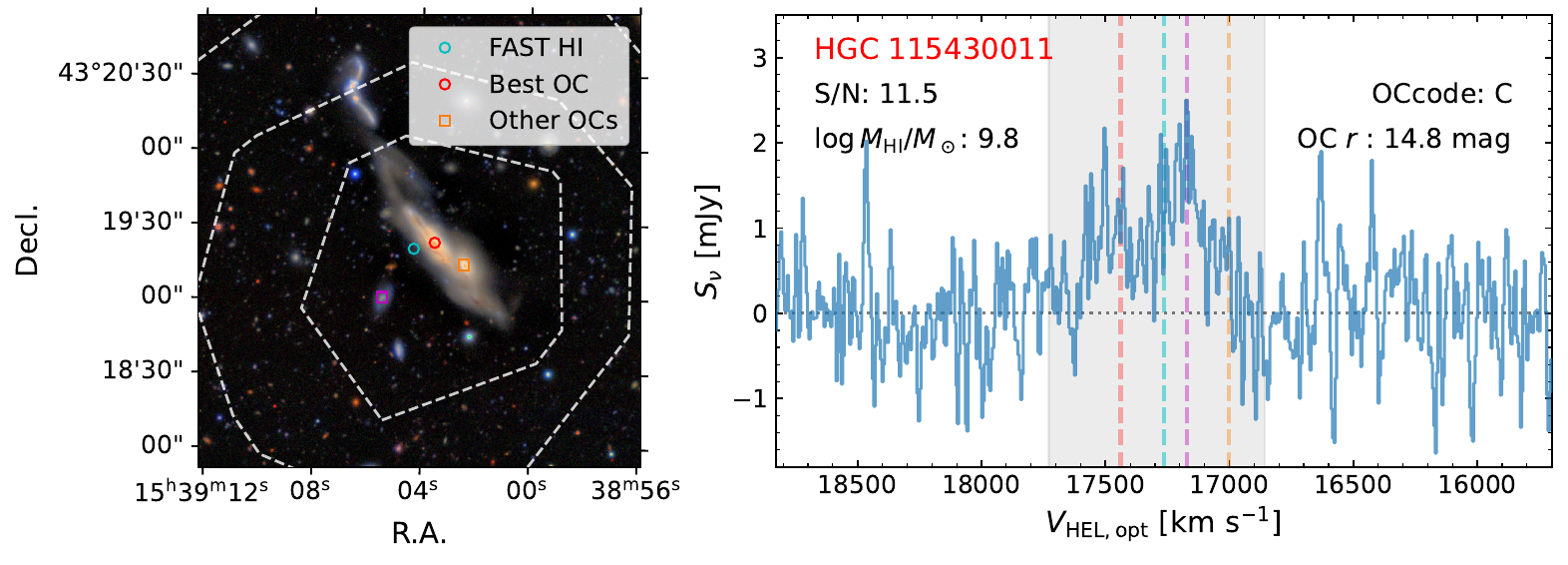}

\includegraphics[width=1.5\columnwidth]{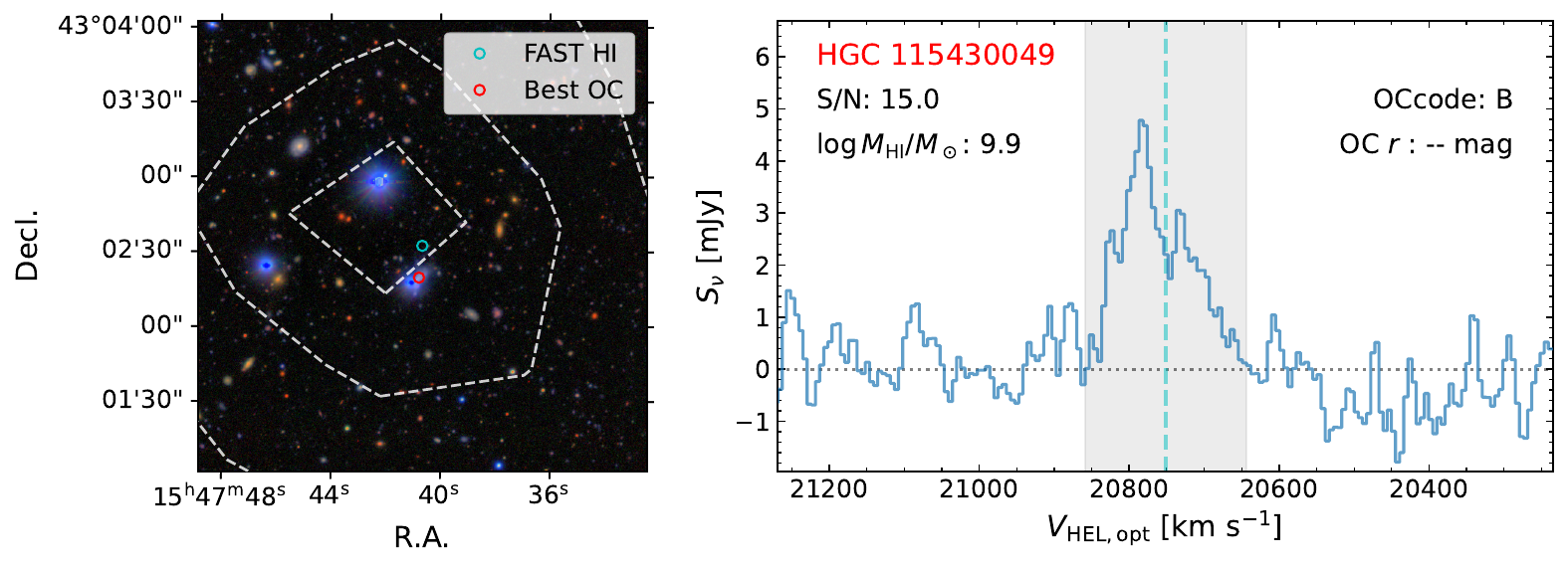}

\includegraphics[width=1.5\columnwidth]{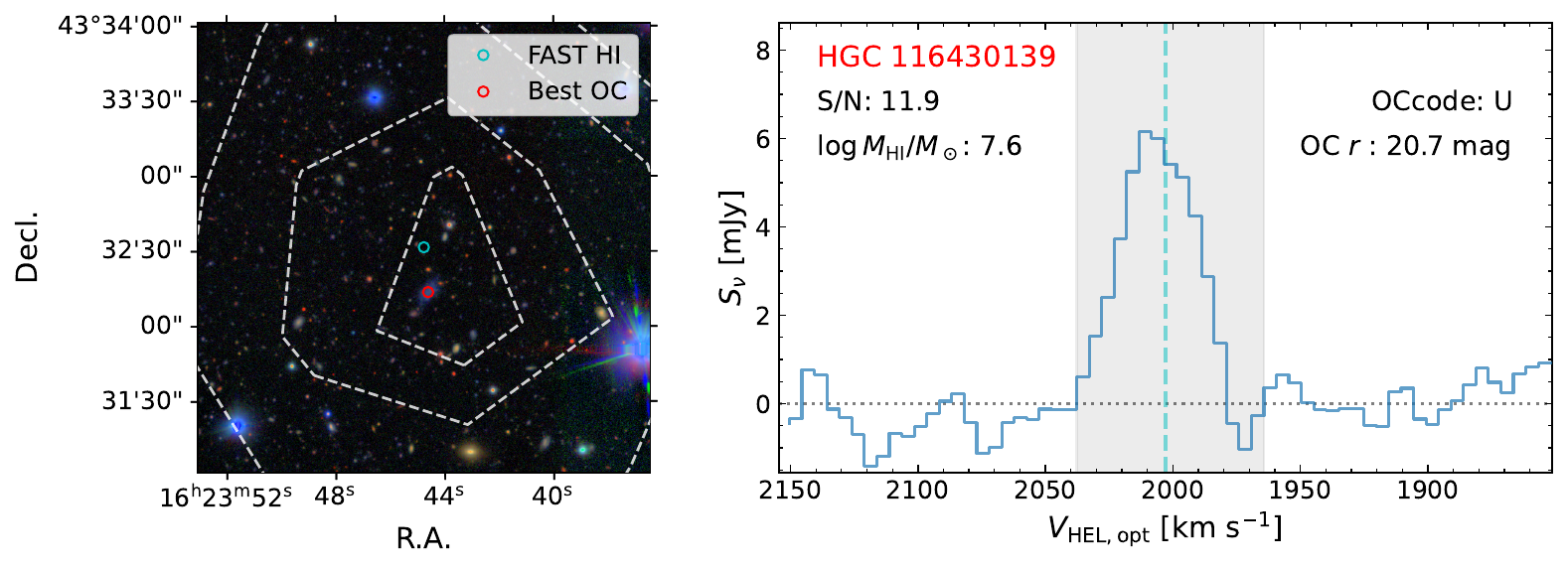}

\includegraphics[width=1.5\columnwidth]{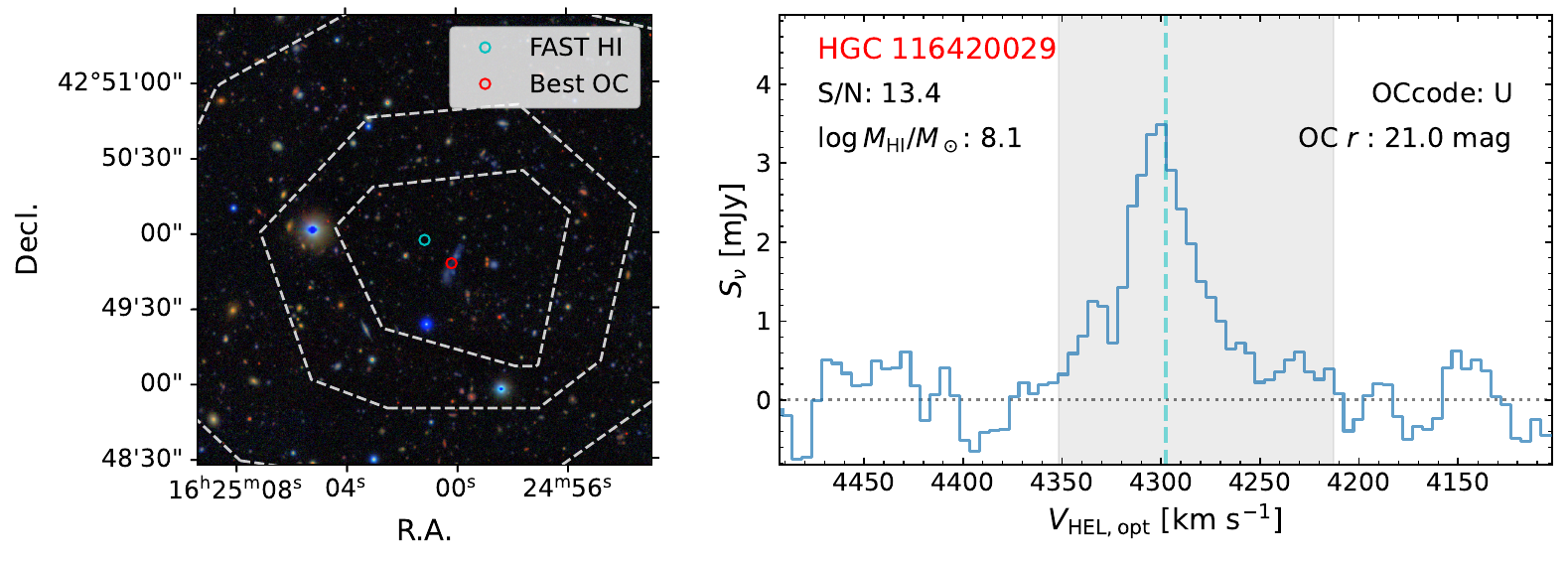}
\caption{Same as Figure~\ref{fig:app_optspec}, but for sources with representative OC cases. The first example shows a case with confusion (OC code \texttt{C}), where the best OC is marked with a red circle and other OCs are indicated by orange and magenta squares. Their corresponding redshifts are also displayed in the same colors.
The second row presents a source whose OC is heavily contaminated by the flux of a nearby foreground star, preventing a reliable $r$-band magnitude measurement (OC code \texttt{B}). 
The bottom two rows show sources with no available spectroscopic redshift, where the associated optical galaxy is too faint for reliable spectroscopy and the photometric redshift is unreliable (OC code \texttt{U}).}
\label{fig:app_optspec3}
\end{figure*}

We present representative \HI\ detections spanning a range of S/N values to illustrate the data quality and robustness of our catalog. Each panel shows the HSC-SSP optical image ($1.5' \times 1.5'$) centered on the \HI\ position overlaid with \HI\ flux contours, alongside the corresponding \HI\ spectrum.
Figures~\ref{fig:app_optspec} and \ref{fig:app_optspec2} show six sources ordered by decreasing S/N, including both \SOFIA-identified detections and a source recovered through targeted search at a known DESI redshift position (HGC~115420003, S/N~$=3.8$).
Figure~\ref{fig:app_optspec3} presents three representative optical counterpart cases: one interacting galaxy pair showing confusion (OC code \texttt{C}), 
one source whose optical photometry is contaminated by a bright foreground star (OC code \texttt{B}), 
and two sources lacking reliable spectroscopic redshifts, because their optical counterparts are too faint for spectroscopy and the photometric redshift is unreliable (OC code \texttt{U}).

\bibliography{ref}{}
\bibliographystyle{aasjournal}

\end{document}